\documentclass[preprint,11pt]{elsarticle}

\usepackage{amssymb}
\usepackage{float}
\usepackage{graphicx}
\usepackage{algorithm}
\usepackage{amssymb}
\usepackage{amsmath}
\usepackage{breqn}
\usepackage{multirow}
\usepackage[math]{cellspace}
\usepackage{array}
\usepackage{listings}
\usepackage{csquotes}
\graphicspath{ {figures/} }
\usepackage{framed,color}
\usepackage{geometry}
\usepackage{tabularx}
\usepackage{subfig}

\usepackage{enumitem}
\definecolor{shadecolor}{rgb}{0.9,0.9,0.9}
\definecolor{myGreen}{rgb}{0.1,0.6,0.1}
\definecolor{mypink}{rgb}{0.6,0.1,0.1}
\definecolor{myBlue}{rgb}{0.1,0.1,1}
\definecolor{myGray}{rgb}{0.8,0.8,0.8}

\journal{Arxiv.org}

\usepackage{hyperref}

\begin{document}
\begin{frontmatter}

\title{Benchmarking the Benchmark - Analysis of Synthetic NIDS Datasets}

\author{Siamak Layeghy, Marcus Gallagher, Marius Portmann}

\address{School of ITEE, The University of Queensland, Brisbane, Australia}

\begin{abstract}





Network Intrusion Detection Systems (NIDSs) are an increasingly important tool for the prevention and mitigation of cyber attacks. Over the past years, a lot of research efforts have aimed at leveraging the increasingly powerful models of Machine Learning (ML) for this purpose. 
%
%
A number of labelled synthetic datasets generated have been generated and made publicly available by researchers, and they have become the benchmarks via which new ML-based NIDS classifiers are being evaluated. 
Recently published results show excellent classification performance with these datasets, increasingly approaching 100 percent performance across key evaluation metrics such as accuracy, F1 score, etc.
Unfortunately, we have not yet seen these excellent academic research results translated into practical NIDS systems with such near-perfect performance.

This motivated our research presented in this paper, where we analyse the statistical properties of the benign traffic in three of the more recent and relevant NIDS datasets, (CIC, UNSW, ...).
As a comparison, we consider two datasets obtained from real-world production networks, one from a university network and one from a medium size Internet Service Provider (ISP).
%
%
Our results show that the two real-world datasets are quite similar among themselves in regards to most of the considered statistical features. Equally, the three synthetic datasets are also relatively similar within their group.
However, and most importantly, our results show a distinct difference of most of the considered statistical features between the three synthetic datasets and the two real-world datasets. 
Since ML relies on the basic assumption of training and test datasets being sampled from the same distribution, this raises the question of how well the performance results of ML-classifiers trained on the considered synthetic datasets can translate and generalise to real-world networks. 
We believe this is an interesting and relevant question which provides motivation for further research in this space.

\end{abstract}

\begin{keyword}
Network traffic, feature distribution, IDS Dataset, NetFlow, Traffic Characteristics
\end{keyword}

\end{frontmatter}

\section{Introduction}\label{introduction}
Network Intrusion Detection Systems (NIDSs) are an important defence mechanism to protect computer networks against an increasing diversity, sophistication and volume of cyber attacks. 
Due to the tremendous progress in Machine Learning (ML) over the last few years, in particular Deep Neural Networks (DNNs), there has been a lot of recent research into leveraging the power of novel ML models for Network Intrusion Detection Systems. 
In particular supervised ML methods have shown great potential over traditional signature-based NIDS.
%
As with any supervised ML problem, the availability of high quality labelled datasets is absolutely critical.
For many years, the KDD99 dataset \cite{KDD99} has been the most widely used benchmark dataset for the evaluation of NIDSs. 
However, it is well established that the dataset has significant limitations~\cite{UNSW-NB15}
the main one being its age, it was created over two decades ago. 
Given that Youtube, Facebook, Spotify, mainstream cloud computing and smartphones did not exist when the dataset was created, one can appreciate that the pattern of network traffic has undergone a profound change since then. 
%


Furthermore, the type and sophistication of network attacks have undergone an equally dramatic change in the last 20+ years 
\footnote{It is quite suprising that the KDD99 dataset is still used today???}.
The need for more recent and relevant NIDS datasets has been clearly identified~\cite{UNSW-NB15}, 
and has lead to the development of a range of new datasets over the last few years. 
In contrast to ML application areas such as image classification, where high quality benchmark datasets can relatively easily be generated, this is a much harder problem in the context of NIDSs. 
Ideally, we would have datasets collected from real production networks, with realistic network patterns of benign traffic, together with a wide range of correctly labelled attack traffic. 
Since such ideal NIDS datasets are not readily available, researchers have recently developed a range of new synthetic datasets, which have become the new benchmarks. 
These synthetic datasets are typically generated in a controlled and relatively small simulation or test-bed environment, where both normal traffic and attack traffic are created and labelled. 
Each of these datasets typically has its own dedicated feature set, which is collected and represented in a flow-based format. 
%
Over the past few years, researchers have extensively used these synthetic datasets to evaluate a wide range of new proposed ML-based intrusion detection models and methodologies. 
Recently published results show increasingly excellent classification performance, approaching 100\% across the key performance metrics, such as accuracy, F1-score, etc. Consequently, one could assume that the problem of ML-based NIDS has been largely solved. 
%
%
%
Arguably, this is not quite the case, and the excellent results achieved in recently published academic research have not yet translated into practical, near-perfect intrusion detection systems deployed in real-world production networks. 
This apparent gap has motivated our research presented in this paper. 

ML generally assumes that statistical properties of training data are the same as for testing data. Therefore, in order for the performance of an ML-classifier trained and evaluated on synthetic datasets to generalise and translate to a real network scenario, the statistical properties of both datasets would have to be similar.
%
Our aim was therefore to compare the statistical properties of synthetic NIDS datasets with those of real world network traffic. 
%
%
Our aim was therefore to compare the statistical properties of synthetic datasets with those of network traffic obtained from real, large-scale production networks. In our analysis, we focused on benign (non-attack) traffic, due to the lack of attack labels in the production network traffic available to us. 
For our analysis, we have considered three recently published and widely used synthetic datasets UNSW-NB15~\cite{UNSW-NB15}, CIC-IDS2017~\cite{CIC} and TON-IOT~\cite{TON-IOT}.
The datasets contain 44 to 85 number of features in different formats. 
We further used two datasets from large-scale production networks, collected in 2019, one from a medium sized Australian ISP, and the other from the University of Queensland network with $\sim$100 and $\sim$700 flows per second respectively. 
The real-world datasets were in Netflow/IPFIX format, which is widely available in  

In order to enable the comparison of the five datasets, we require them to be in the same format. 
For this, we leveraged our previous work \cite{netflow_datasets}, where we converted synthetic NIDS datasets from their proprietary feature set and format to the standard Netflow/IPFIX format \footnote{Datasets are available here ...}.  In \cite{sarhan2021standard}, we argue the benefits of having a standard feature set and dataset format, and also show that, somewhat surprisingly, ML-classifiers achieve higher performance using the Netflow feature set and format, compared to the original version.  

For our  comparison, we considered 9 practically relevant statistical features, such as flow duration, flow size, packet size, plus a number of IP address and port number related features. 
%
We compared the statistical distributions via box plots and CDF of each feature across the five considered datasets. We further quantified the distance of the different feature distributions using the Wasserstein metric~\cite{Ramdas2017}.
Finally, we calculated and visualised the embedding of the 9 features into a 2 dimensional feature space, using four different embedding algorithms.

Our analysis provided the following key findings. The two real-world datasets, despite the fact that they had been collected from quite different networks, exhibit a high degree of similarity in the traffic feature distributions. Similarly, the synthetic datasets are quite similar amongst themselves in regards to most traffic features.  
However, and most interestingly, our analysis found a highly significant difference between the synthetic datasets and the real-world datasets in regards to most of the considered feature statistics. 

To the best of our knowledge, this paper provides the first analysis of recent synthetic NIDS datasets and their comparison to real word traffic, 
We believe our results are relevant due to the extensive use of these synthetic datasets as a benchmark to evaluate ML-based NIDS models and algorithms, and they motivate future research into the development of new datasets that more closely  match the properties of traffic in large scale real-world networks. 
This is an important goal in order to allow the translation and generalisation of the excellent NIDS performance achieved in academic research into  NIDSs that are increasingly practically relevant and widely deployed in practical settings.

The rest of this paper is organised as follows. After explaining the background in the next section, the three synthetic datasets
along with the two real-world network traffic datasets that are used as references for this purpose, are introduced in Section \ref{Datasets}. The next section provides various qualitative comparisons between features  distributions of the synthetic datasets and the real-world   datasets. Then, Section \ref{Analysis} quantifies the differences of feature distributions among two groups of datasets. Section \ref{Related Works} provides the related works in the field and Section \ref{Conclusion} concludes the paper and provides the insights from our study.

\section{Background}\label{Background}

\subsection{Network Intrusion Detection Systems}
The main approaches for building network-based intrusion detection systems (IDS) include \textit{misuse detection} or \textit{rule-based}, \textit{anomaly detection}, and a third \textit{hybrid} approach, which tries to combine benefits of the first two approaches \cite{MonowarH.Bhuyan2014}. In misuse detection systems, detection is based on the known attack signatures or rules. In anomaly detection, most of the methods define a normal traffic model and then identify anomalies as deviations from the normal model. While misuse-based IDSs have low false positive ratios, they fail to detect new types of attacks / intrusions since they have designed to detect specific types of intrusions. Anomaly-based IDSs, on the other hand, have higher false positive ratios but are potentially capable of detecting unknown attacks / intrusions. The anomaly-detection systems are implemented using various techniques, and hence there are various categories of anomaly-based IDS \cite{Hung-JenLiao2013}.

The main approaches for the network anomaly detection include \textit{Statistical}, \textit{Machine Learning}, \textit{Soft Computing}, \textit{Knowledge-based}, and \textit{Combination Learner} methods. In the statistical methods, generally a model is fitted to the given data (usually normal behaviour), either via parametric or non-parametric techniques. Then, by applying statistical inference tests, the probability that an instance is generated by the model is calculated. If such probability is low, the instance is considered anomaly \cite{MonowarH.Bhuyan2017}. 
In the soft computing approach, which includes methods based on Genetic Algorithms (GA), Fuzzy Set Theoretic (FST), Rough Set (RS), and Ant Colony and Artificial Immune Systems (AIS), persistant features of data are detected and categorized without environmental feedback. For instance, in methods based on GA, heuristic search techniques based on evolutionary ideas are used to learn the user profiles, and in methods based on FST, fuzzy rules are used to determine the likelihood of specific or general network attacks \cite{MonowarH.Bhuyan2014}.
In knowledge-based methods, network events are matched against predefined rules or patterns of attacks, i.e. attack signatures. Several approaches have been taken in the knowledge-based methods including expert systems, rule-based, ontology-based, logic-based and state-transition analysis \cite{MonowarH.Bhuyan2017}.
 The main ingredient of all combination learner methods, which include  ensemble-based, fusion and hybrid approaches, 
 is combining advantages of different methods to enhance the performance of anomaly detection. In the case of ensemble-base and fusion approaches various classifiers are combined and the hybrid approach enhances the anomaly detection by getting the advantages of misuse detection methods~\cite{MonowarH.Bhuyan2014}.
 
Among different methods utilised for implementing the anomaly-detection IDS, systems based on \textit{machine learning} have been very common~\cite{kumar}. This approach includes two methods clustering / outlier-based, and classification-based.
While clustering-based methods do not require labelled datasets for their training / operation, their evaluation necessitates using labelled datasets.
Training and evaluation of the second sub-category of machine learning based approach, i.e. classification-based methods, require labelled datasets in which classes / categories of data are known and included in the dataset as \textit{labels}. This allows machine learning algorithms to learn the patterns of various classes and accordingly classify the dataset records. In the field of anomaly detection, these classes are mainly \textit{normal} and \textit{anomaly} classes. The anomaly class can be further divided into different types of anomalies such as network failures, intrusions and other types of attacks.

\section{Datasets Explored in This Study}\label{Datasets}
In this study we have used three Synthetic / testbed-based IDS benchmark datasets which we are compared with two real-world network traffic records.

\subsection{Synthetic Datasets}
The Synthetic / testbed-based datasets selected for this study are chosen from the most recent IDS benchmark datasets that published from 2015 till 2019.

\subsubsection{UNSW-NB15 Dataset}
This is the oldest dataset among the three selected datasets, which is published in 2015  by the Cyber Range Lab of the Australian Centre for Cyber Security (ACCS)~\cite{UNSW-NB15}. 
The dataset has consisted of 2,540,044 flows including 87.35\% benign and 12.65\% attack flows. The traffic flows are represented in 49 features generated by Argus and Bro-IDS from traffic of a test bed in which 9 different attack types are combined with normal background testbed traffic. This dataset is also published in the form of raw pcap files.

\subsubsection{CIC-IDS2017 Dataset}
This dataset has been generated by Canadian Institute for Cybersecurity in 2017 \cite{CIC}. The main priority in generating this dataset is stated by authors as having realistic background traffic. They have used their own developed tool for network analysis, CICFlowMeter, to generate a dataset in which flows are labeled based on time stamp, source, and destination IPs, source and destination ports, protocols and attack. They have also provided the dataset represented by extracted features definition. 
They have included 21 different network attack in this dataset, which includes most of the major network breaches such as DoS, DDoS, different scans, Botnet, and various types of Web and Infiltration attacks. They ran individual class of attacks separately and provided them in a separate CSV files, as for normal background traffic of their test bed. The dataset is also published in raw pcap format.

\subsubsection{TON-IOT Dataset}
This is a newer dataset published by ACCS in 2019, encompassing a broader scope  compared to their previous dataset~\cite{UNSW-NB15}, which includes  traffic records from their testbed network and IoT devices along with logs of the operating systems~\cite{TON-IOT}. The network traffic records, explored in our work, are represented in the form of 22,339,021 network flows using 44 features extracted by Bro-IDS. The background / benign traffic constitutes 3.56\% of dataset and the remaining flows (96.44\%) are 9 different attack records. This dataset is also published in the raw pcap format.

\subsection{Real-World Network Traffic Records}
In order to compare the above testbed-based datasets with real-world traffic, we collected two sets of network traffics. Since we are looking for the aspects of network traffic that must be valid in any network type, we selected two completely different networks. This is an important basis in our study to avoid biased judgement and be able to generalise the results to any network type. As such, the two network traffic records explained in this section, not only are collected from two different types of networks, but they also belong to two different organization with various administration requirements and ruling polices who serve different types of customers. In addition, these two sets of network traffic have been recorded in two different time period, 2017 and 2019, to maximise the possibility of result generalisation.

\subsubsection{Internet Service Provider Traffic Dataset}
The first real-world dataset utilised in this study is the NetFlow records from an Australian Internet Service Provider (ISP) backbone network. This ISP is a provider of enterprise-grade managed services such as Internet, MPLS, VoIP, and cloud with dual point of presences (PoPs) in all major Australian capital cities, New Zealand, Hong Kong and Manilla. Most of the customers of the ISP are businesses and companies with multiple offices, sometimes thousands of offices around the country, that use ISP's services for the connectivity and other network and Internet services. 
For collecting the traffic records we used the nprobe \cite{Ntopng2017} software on a server to extract NetFlow records with about 20 fields. The whole traffic from ISP's backbone network was mirrored and aggregated toward the nprobe server where after extracting NetFlow records stored them in another collector server. These records reflect the whole traffic in the monitored part of ISP's backbone network without any sampling or exemptions. We have collected these NetFlow records for about 30 days in June 2017 and it includes about 400GB of flow records.

\subsubsection{University of Queensland (UQ) Dataset}
The next real-world dataset we have used in this study includes NetFlow records collected form the LAN of faculty of Engineering, Architecture and Information Technology (EAIT), The University of Queensland. The EAIT faculty consists of five schools including Architecture, Chemical Engineering, Civil Engineering,    Information Technology and Electrical Engineering, and Mechanical and Mining Engineering. 
We used similar setup for collecting this dataset, i.e. using nprobe NetFlow exporter with the same set of fields that we used on ISP NetFlows.
The data collection in this site started on early Feb. 2019 and continued for about 50 days. 
The recorded data includes all the traffic flows in the monitored part of the network including all wired and wireless communications, servers' traffic and all workstations in all subsidiary schools of EAIT faculty, totally about 4TB.

Table~\ref{tab: datasets} shows a summary information relating the five datasets studied in this paper. 
This information includes the type of dataset, i.e. is it a synthetic or real-world dataset, the percentage of attack and benign flows from total records of the dataset, number of features, the year dataset is collected / published, size, type and the tools used to generate / collect the dataset.

\begin{table}[!t]
\tiny
\caption{Summary information of datasets studied in this paper}

\label{tab: datasets}
\centering
\begin{tabularx}{0.85\columnwidth}{ 
|m{2.2cm}
|>{\centering\arraybackslash}X
|>{\centering\arraybackslash}X
|>{\centering\arraybackslash}X
|m{0.8cm}
|>{\centering\arraybackslash}X
|>{\centering\arraybackslash}X|
}

\hline
&&&&&&\\
\centering{\textbf{Dataset}}  & 
\textbf{Synthetic / Real-world}  & 
\textbf{Attack~  (\%)} & 
\textbf{Number of Features} & 
\centering\textbf{Year} & 
\textbf{Format} & 
\textbf{Generation Tool} \\ &&&&&& 

\\  \hline
&&&&&& \\

\centering\textbf{UNSW-NB15}~\cite{UNSW-NB15} & Synthetic & 12.65 & 49 & \centering2015 & proprietary flow format  & Argus / Bro\\ &&&&&& \\
 \hline
&&&&&& \\

\centering\textbf{CIC-IDS2017}~\cite{CIC} & Synthetic & 28.16 & 85 & \centering2017 &  proprietary flow format & CICFlowMeter \\ &&&&&&\\ 
\hline
&&&&&& \\

\centering\textbf{TON\_IOT}~\cite{TON-IOT} & Synthetic & 96.44 & 44 & \centering2019 & proprietary flow format & Security Onion / Bro \\ &&&&&& \\
\hline
&&&&&& \\

\centering{\textbf{ISP}~[Our]} & Real-World & 0 & 20 &  \centering2017 & NetFlow & nprobe \\ &&&&&&
\\ \hline
&&&&&& \\

\centering{\textbf{UQ}~[Our]} & Real-World & 0 & 20 &  \centering2019 & NetFlow & nprobe \\ &&&&&&
\\ \hline

\end{tabularx}
\end{table}

\section{Statistical Analysis of Network Traffic Features}\label{Comparison}

As discussed in Section \ref{NAC}, anomalies typically change statistical distribution of some traffic features~\cite{Lakhina, Soule2007, histo_anomaly}. 
Hence, we cannot expect to acquire similar results from the comparison of  statistical distributions of these benchmark datasets containing anomalies / attacks with that of  the real-world traffic records with unknown attack inclusion status. So, we needed to make sure that the selected parts of both groups of datasets are attack free (normal traffic).

As such, we had two issues to address before starting our analysis. First, to select attack free parts from benchmark datasets, and second, to make sure the selected parts of the real-world traffic does not include attacks. 
The first problem was relatively easy, as normal / background traffic were provided in separate files in UNSW-NB15~\cite{UNSW-NB15}, CIC-IDS2017~\cite{CIC} and TON\_IOT~\cite{TON-IOT}, and we used the provided metadata to exclude all the attack flows. 
The second task was rather difficult. Initially, we asked for the attack / anomaly logs in the ISP and our university IT department related to the selected part of the traffic. 
Then, we applied our under-development AI-based anomaly detection algorithms on these recordings. Finally, any detected anomalies by algorithms were investigated in complete details and discussed with the corresponding network administration for final decision. In this way, an anomaly-free part of the traffic was selected equivalent to 24 hours of both ISP and UQ traffic records. 

Table~\ref{tab: features} shows the list of traffic features we have used in this statistical distributions comparison. The first column is the name of feature, the second column is the NetFlow (V9) fields used to calculate the feature, and the third column is the formula / equation applied for the calculation. 
Since some of the needed NetFlow fields were not provided in the original published benchmark datasets,  we had to use their provided packet captures (pcap files) to generate the NetFlow records. For this, after determining the attack free parts of the benchmark datasets, as explained above, the corresponding pcap files were identified and fed to nprobe \cite{Ntopng2017} software to generate NetFlow records.

All the NetFlow fields utilised in our study have previously been used for characterising anomaly traffic in at least one or more studies~\cite{Lakhina, Soule2007, histo_anomaly}, except the \texttt{L7\_PROTO} that was not easily provided by devices in the past. As such, comparing these datasets with real-world traffic, in terms of statistical distributions of these features, provides a realistic measure of similarity to the real-world traffic. 
This is a missing meaningful benchmarking of these datasets that can indicate their suitability for benchmarking the IDS algorithms and systems in the real-world scenarios.

In this study we provide two sets of comparisons. Initially we qualitatively investigate distribution of these features by comparing their \textit{boxplots} and \textit{Cumulative Distribution Functions (CDFs)}, and later their embeddings, in this section. 
This enables us to observe the difference in distribution of these features between the real-world traffic and testbed-based datasets. 
Then we quantify these qualitative comparisons using quantitative distance metrics in the next section.

\begin{table}[!t]

\caption{List of Traffic features investigated in this study, along with NetFlow fields utilised to calculate these features and how they are calculated}

\label{tab: features}
\resizebox{\columnwidth}{5cm}{
\begin{tabularx}{\columnwidth}{ | p{6.4cm} | X | p{3.4cm} | }
\hline

\textbf{Feature}  & \textbf{NetFlow Fields}  & \textbf{How to }  \textbf{Calculate}  \\  \hline

\texttt{flow duration} & \texttt{FIRST\_SWITCHED(FS)},  \hspace{2cm} \texttt{LAST\_SWITCHED(LS)} & \texttt{Ls - FS} \\ \hline

\texttt{flow size (in Bytes)}  & \texttt{IN\_BYTES(IB)}, \hspace{2cm} \texttt{OUT\_BYTES(OB)} & \texttt{IB + OB} \\ \hline

\texttt{packet time (average)} & \texttt{FIRST\_SWITCHED(FS)}, \texttt{IN\_PKTS(IP)}, \texttt{LAST\_SWITCHED(LS)},  \texttt{OUT\_PKTS(OP)} & \[\textstyle\frac{\text{\texttt{LS - FS}}}{\text{\texttt{IP + OP}}} \hspace{3cm} \] \\  \hline

\rule{0pt}{10pt}
\texttt{packet size} & \texttt{IN\_BYTES(IB)}, \hspace{4cm} \texttt{OUT\_BYTES(OB)}, \hspace{4cm} \texttt{IN\_PKTS(IP)}, \hspace{4cm} \texttt{OUT\_PKTS(OP)}& \[\textstyle\frac{\text{\texttt{IB + OB}}}{\text{\texttt{IP + OP}}} \hspace{3cm} \] \\ \hline

\texttt{number of source IPs per destination IP} & \texttt{SRC\_IP}, \texttt{DST\_IP} & \texttt{COUNT(SRC\_IP)} \hspace{4cm} per \texttt{DST\_IP} \\ \hline

\texttt{number of source IPs per destination PORT} & \texttt{SRC\_IP}, \texttt{DST\_PORT} &  \texttt{COUNT(SRC\_IP)} \hspace{4cm} per \texttt{DST\_PORT}\\ \hline



\texttt{number of destination IPs per source PORT} & \texttt{DST\_IP}, \texttt{SRC\_PORT}&  \texttt{COUNT(DST\_IP)} \hspace{4cm} per \texttt{SRC\_PORT}\\ \hline

\texttt{number of destination Ports per source port} & \texttt{SRC\_PORT}, \texttt{DST\_PORT}&  \texttt{COUNT(DST\_PORT)} \hspace{4cm} per \texttt{SRC\_PORT} \\ \hline

\texttt{number of L7 protocols per destination port} & \texttt{L7_PROTO}, \texttt{DST\_PORT} & \texttt{COUNT(L7_PROTO)} \hspace{4cm} per \texttt{DST\_PORT} \\ \hline

\end{tabularx}
}
\end{table}

\subsection{Qualitative Comparison of Feature Distributions} \label{Basic Statistical Analysis}
In this section, we are providing boxplot and CDF of all the features listed in Table~\ref{tab: features} for all five datasets, including the three testbed-based datasets and two real-world traffic records, side by side.

\begin{figure}[!b]
    \centering
    \subfloat[\centering ]
    {\includegraphics[width=0.5\columnwidth, height=5cm]{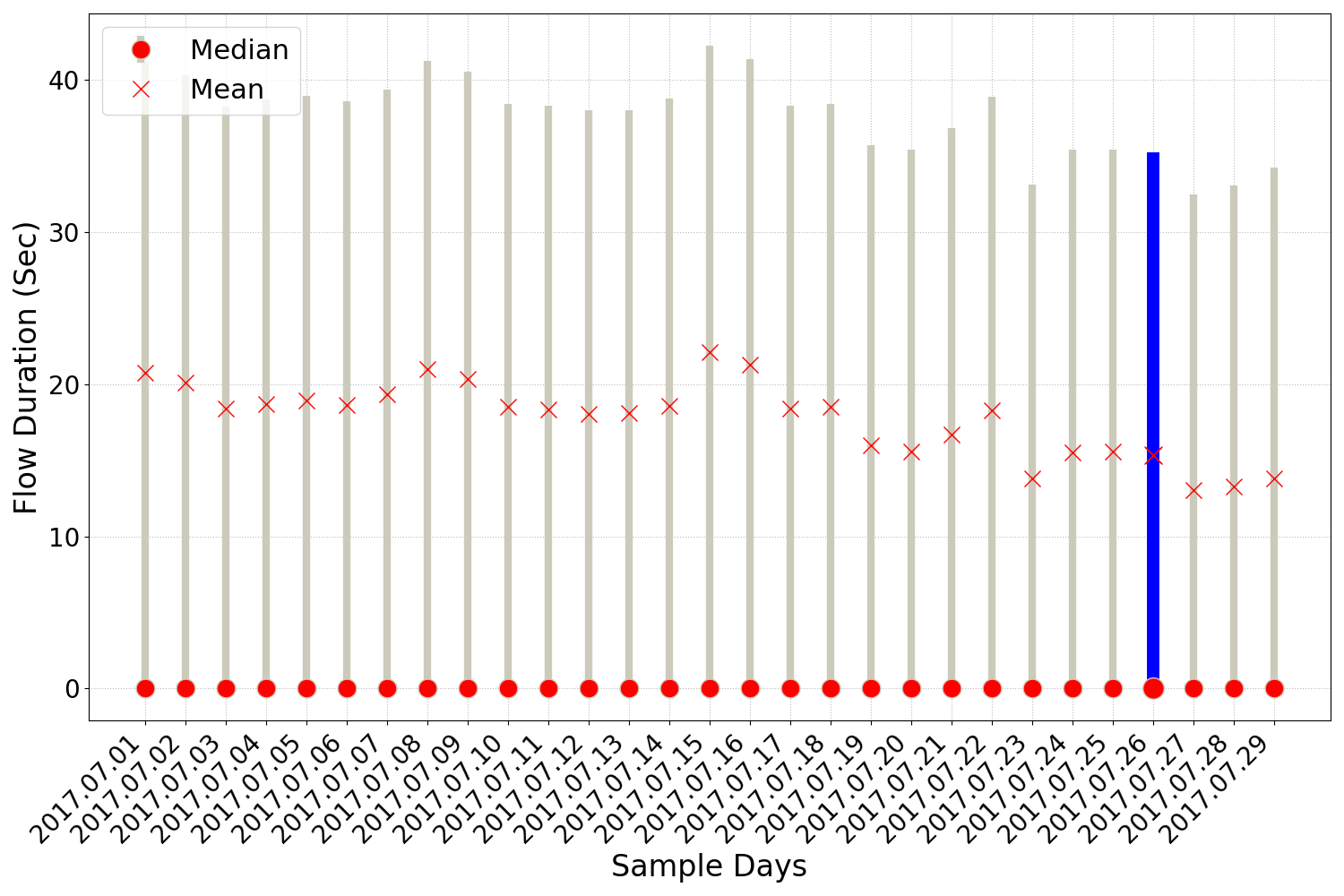}}%
    \subfloat[\centering ]
    {\includegraphics[width=0.5\columnwidth, height=5cm]{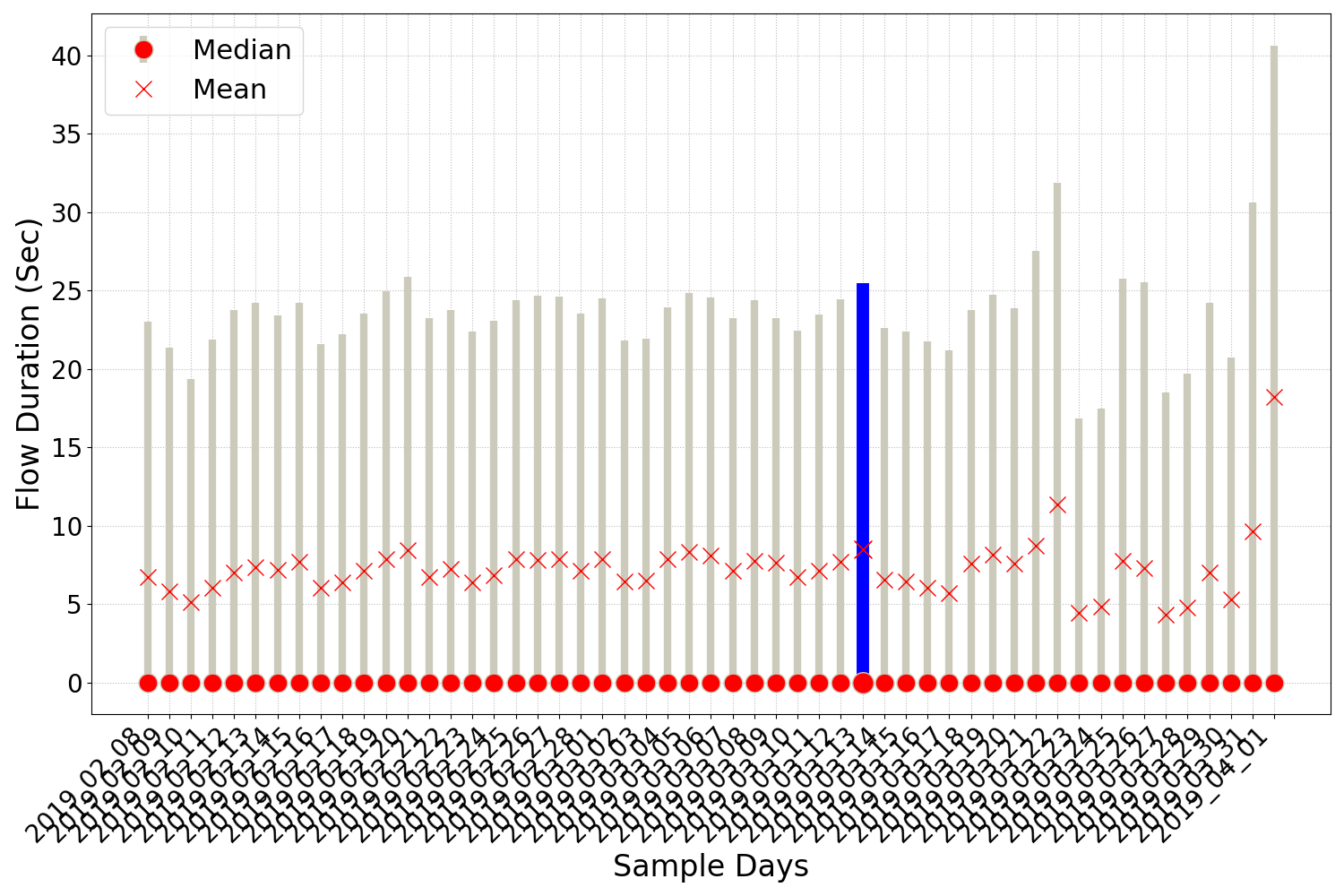}}%
    \qquad
        \subfloat[\centering ]
    {\includegraphics[width=0.5\columnwidth, height=5cm]{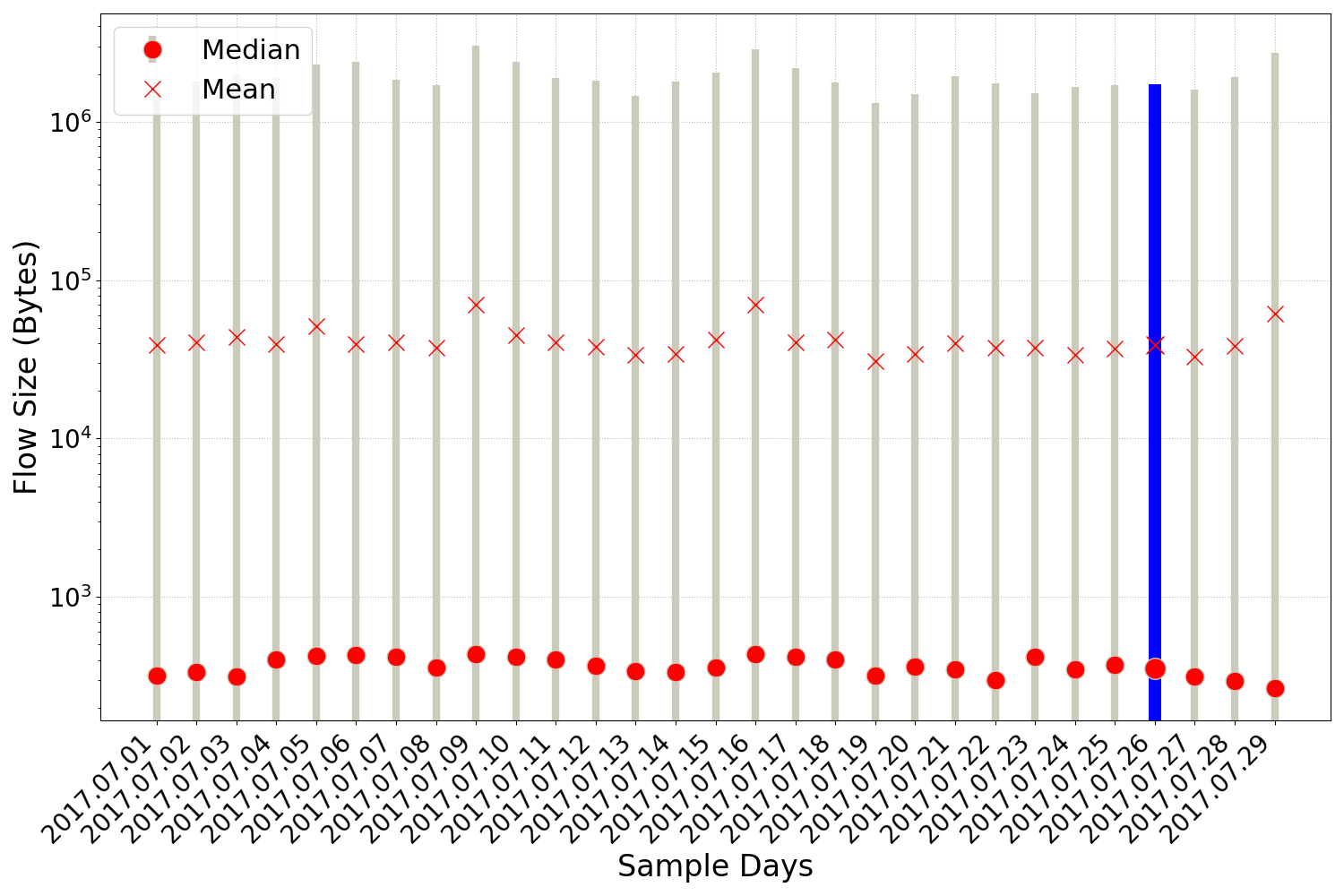}}%
    \subfloat[\centering ]
    {\includegraphics[width=0.5\columnwidth, height=5cm]{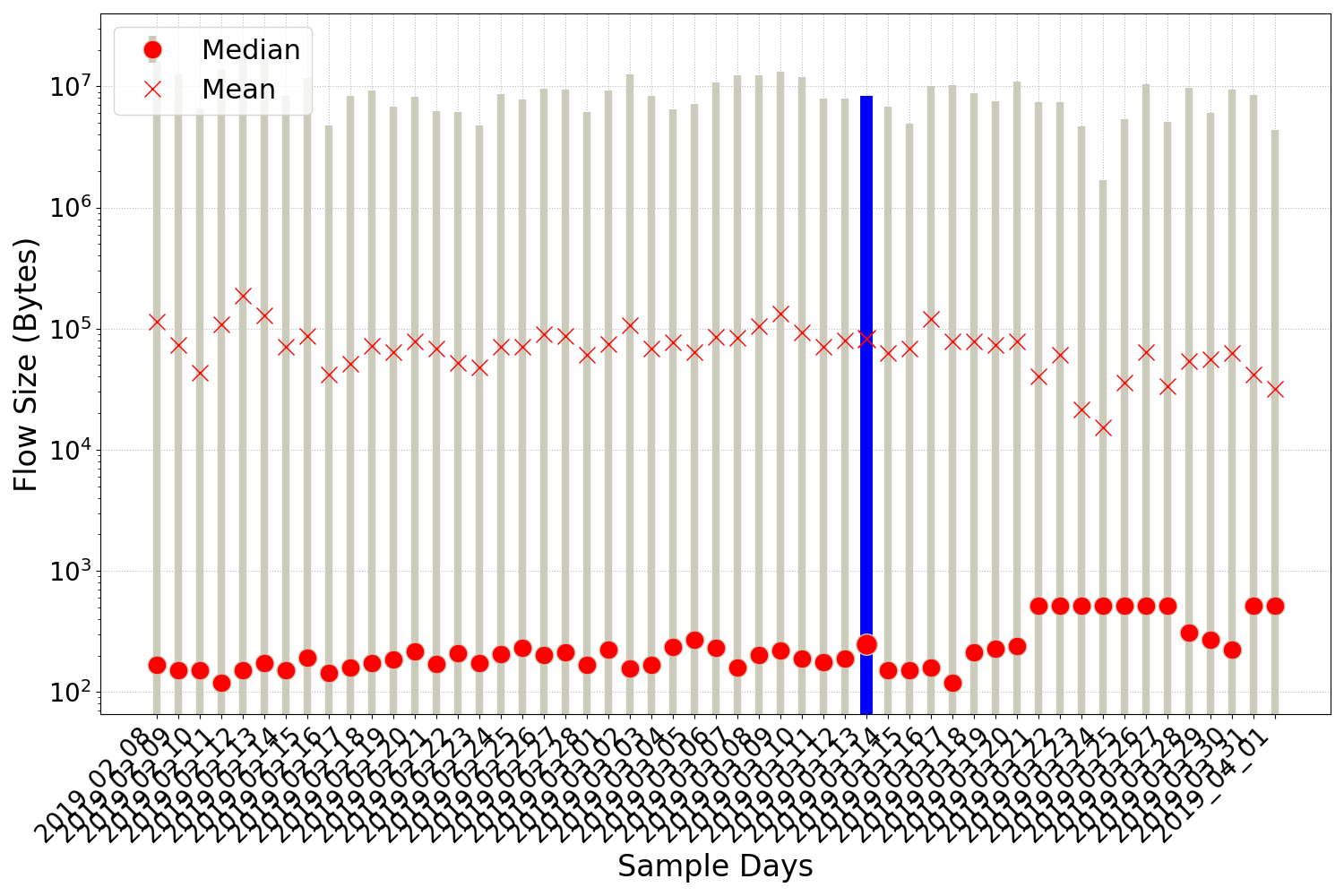}}%
    \caption{Comparing the flow durations and flow sizes within the two real world datasets.  a) and b) show the flow durations of ISP, and sampled UQ dataset, and c) and d) show their flow sizes correspondingly. In each figure, the selected sample day is highlighted in blue color}%
    \label{within datasets}
\end{figure}

\subsubsection{Feature Variability Within the Dataset}
Before comparing the feature distributions between groups, we investigated the feature distribution within various samples of the same dataset. 
Figure \ref{within datasets} shows two examples of comparing feature distribution within the individual datasets. Figure \ref{within datasets}-a and b show the distribution of \textit{flow durations} for different days of ISP and sampled UQ datasets ($\sim20$ sample per minute) respectively, and Figure \ref{within datasets}-c and d show distribution of \textit{flow sizes} for different days of ISP and sampled UQ datasets respectively. In all figures, the red circles show the median, the red crosses show the mean, and the lines / bars show the standard deviation (STD) of the feature. 
The y axis is shown in logarithmic scale for the flow sizes (\ref{within datasets}-c and d) due to the presence of very large flows, which has compressed the main part of the feature range.

As seen, while there are variations in the distribution of features among different days of both datasets, the main characteristics of distributions are similar, i.e. means, and medians are very close and STDs are almost in the same range. This clearly indicates that distribution of these features is an inherent quality of these datasets and its comparison is indeed meaningful. In addition, the it shows that the statistical parameters of selected sample day from each dataset, which is highlighted in blue, is not significantly different from the rest of the dataset.

\subsubsection{Flow Duration}
\begin{figure}[t]
    \centering
    \subfloat[\centering ]
    {\includegraphics[width=0.45\columnwidth, height=5cm]{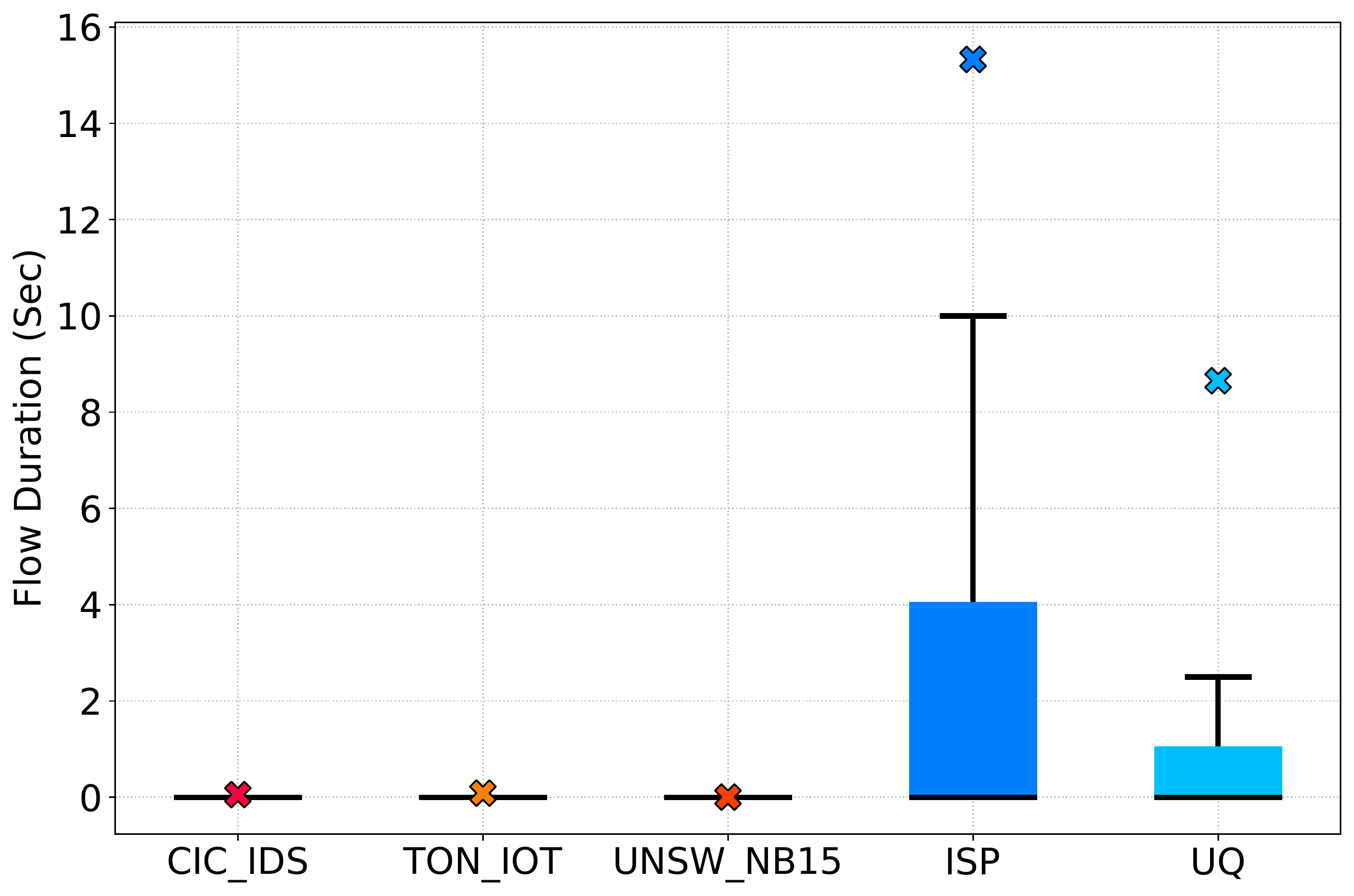} }%
    \qquad
    \subfloat[\centering ]
    {\includegraphics[width=0.45\columnwidth, height=5cm]{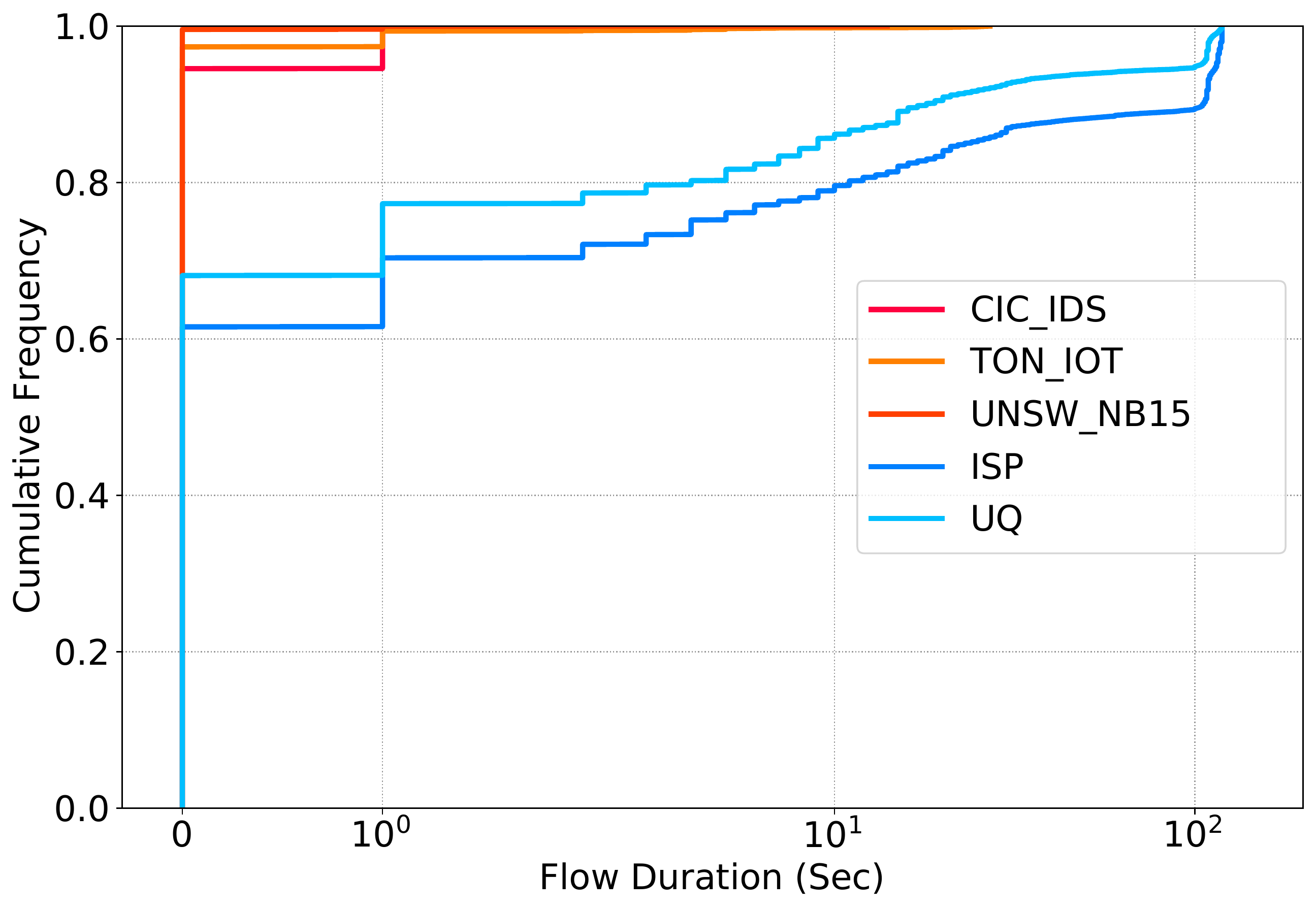}}%
    \caption{Comparing the flow duration in five datasets a) boxplots, and b) CDFs}%
    \label{flow duration}
\end{figure}

Figure \ref{flow duration} shows the distribution of flow duration for all five datasets. In Figure \ref{flow duration}-a  the three

 \noindent left-most boxplots are the three synthetic datasets and two right-most are the real-world traffic records. 
 In all boxplots shown in this paper, the \textit{whiskers} indicate a maximum distance of 1.5 times IQR above and below IQR.
As seen, while the two real-world datasets have completely different IQR from all three synthetic datasets, they have significant overlap. This can be further confirmed by Figure \ref{flow duration}-b in which the CDFs of two real-world traffic records move close together and far from the three synthetic datasets all the way to the right corner of the figure.

\subsubsection{Flow Size (in Bytes)}
\begin{figure}[!t]
    \centering
    \subfloat[\centering ]
    {\includegraphics[width=0.45\columnwidth, height=5cm]{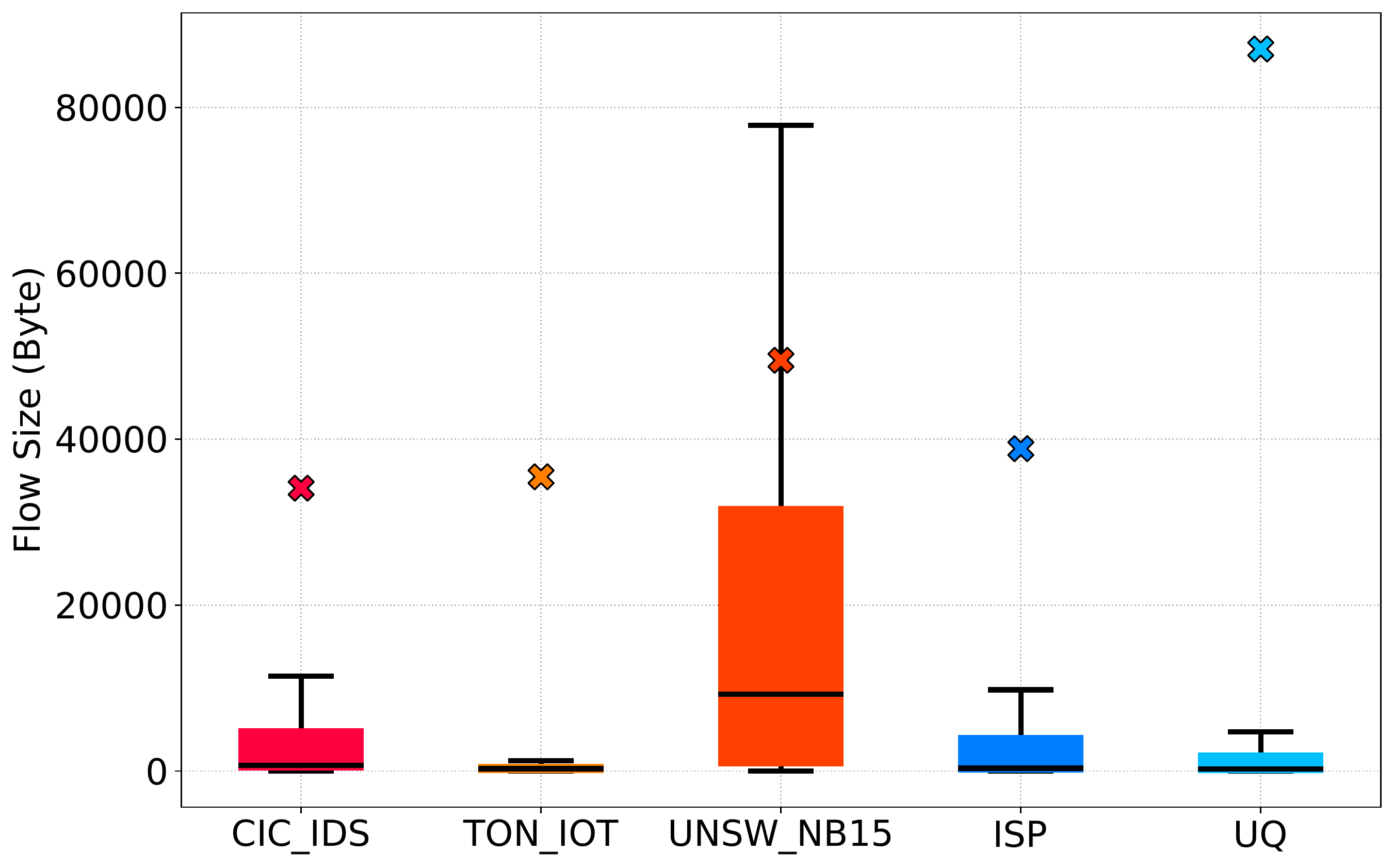} }%
    \qquad
    \subfloat[\centering ]
    {\includegraphics[width=0.45\columnwidth, height=5cm]{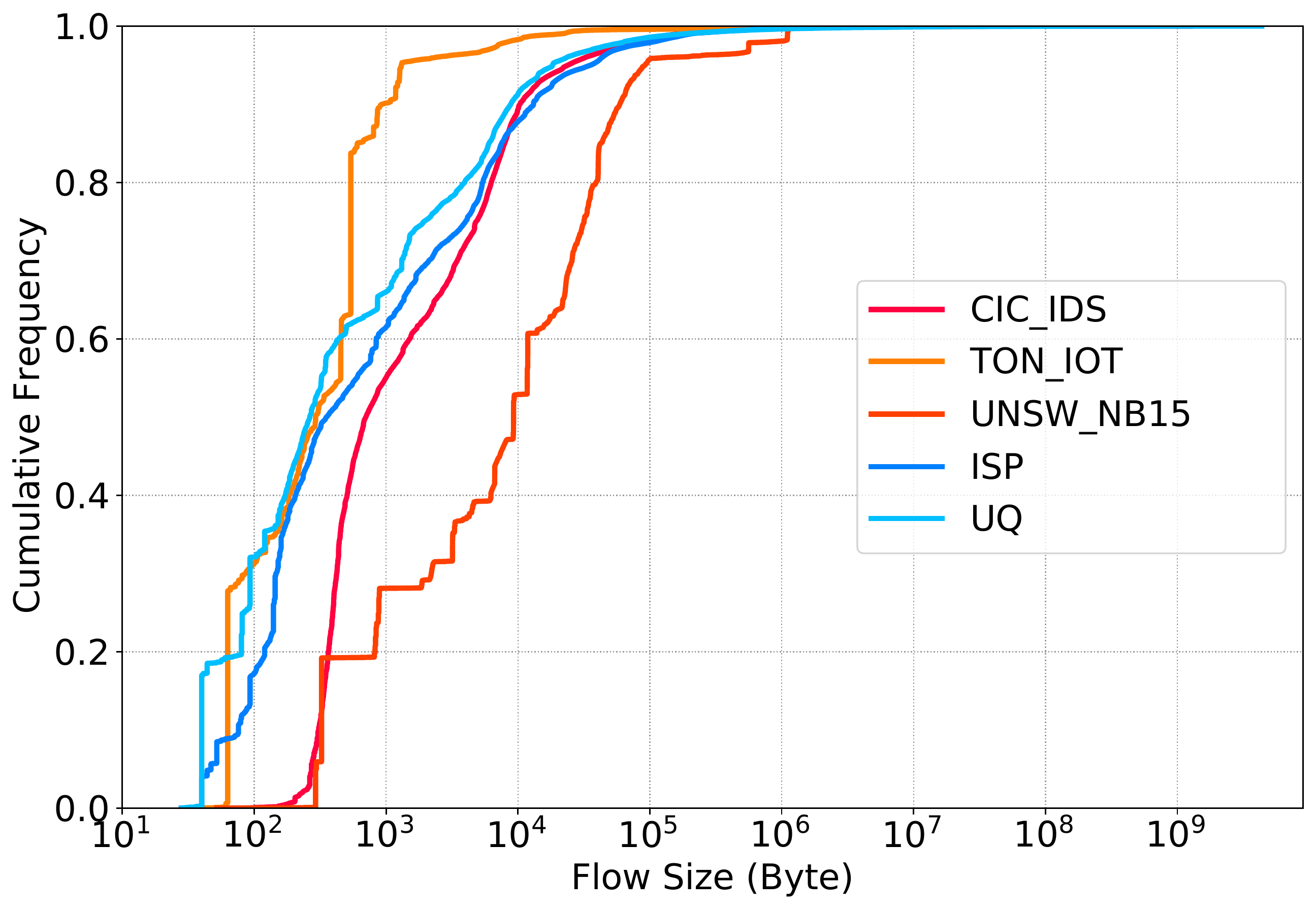} }%
    \caption{Comparing the flow sizes (in Bytes) in five datasets a) boxplots, and b) CDFs}%
    \label{flow size}
\end{figure}

The next feature distribution compared between these five dataset is the flow size (in Bytes). Figure \ref{flow size} compares the distribution of flow sizes in (a) by boxplots and in (b) by CDFs. While in Figure \ref{flow size}-a there is a meaningful distinction between IQRs of  the two synthetic datasets TON\_IOT and UNSW\_NB15, and two real-world datasets, the IQR of CIC has a major overlap with two real-world datasets. In addition, the overlap between IQRs of the two real-world datasets is still significant. This can be further confirmed by investigating the CDF of flow size in these datasets as shown in Figure \ref{flow size}-b. Here, we have used logarithmic scale for the horizontal axis. This is because of the very large outlier values that make the main part of the curves very tiny. The conclusion from figure \ref{flow size}-a can be confirmed here as well. The two real-world datasets close together from the beginning to the end, CIC also closely moves together with them, but TON\_IOT and UNSW\_NB15 have separate path. It worth to mention that we have also computed similar graphs for the distribution of flow sizes in number of packets, i.e. the total number of packets in flows \texttt{(IP + OP)}, but since the result were very similar to flow sizes in Bytes (Figure \ref{flow size}) we gave up to include them.

\subsubsection{Packet Time (average)}
\begin{figure}[!t]
    \centering
    \subfloat[\centering ]
    {\includegraphics[width=0.45\columnwidth, height=5cm]{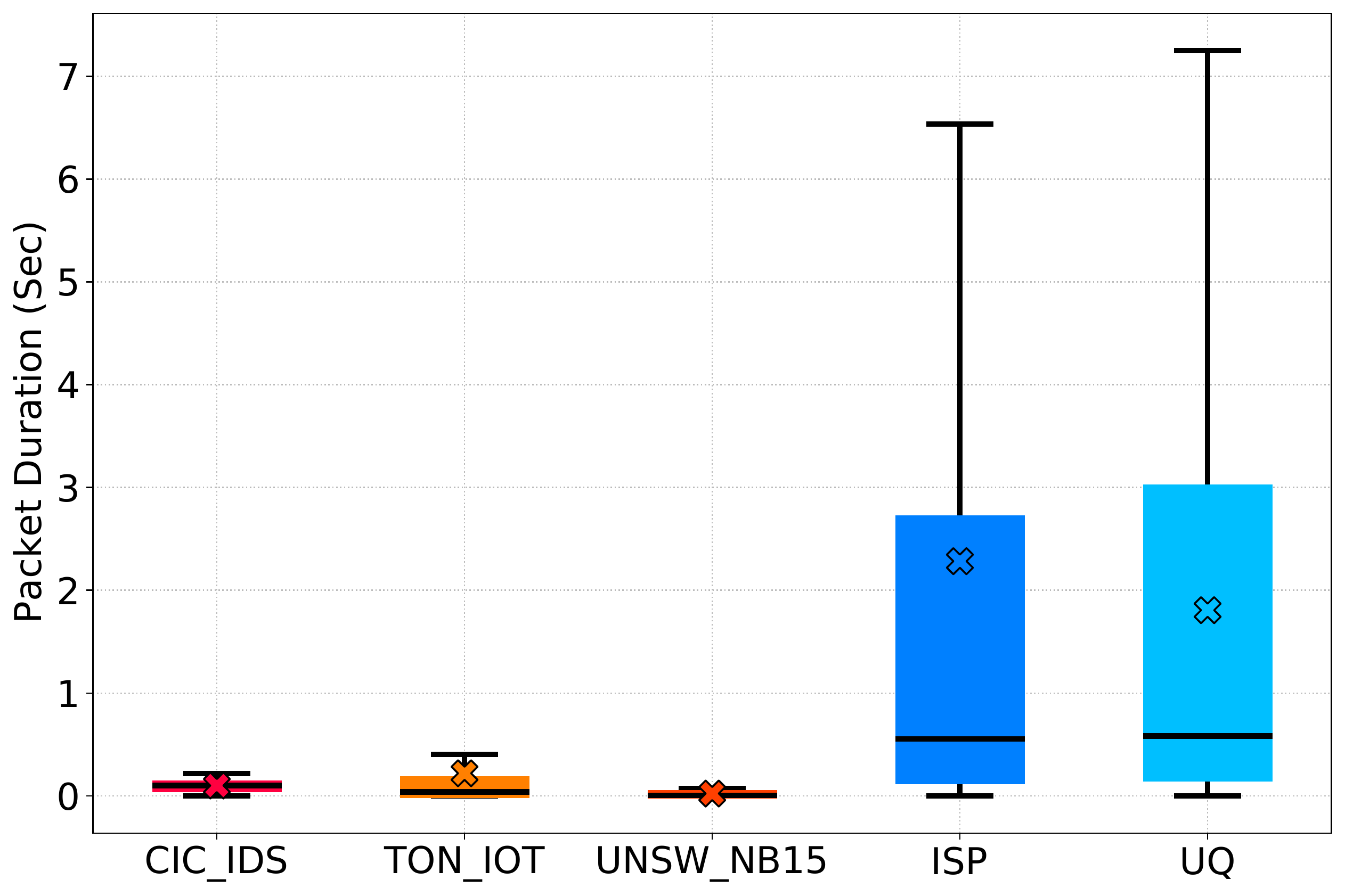} }%
    \qquad
    \subfloat[\centering ]
    {\includegraphics[width=0.45\columnwidth, height=5cm]{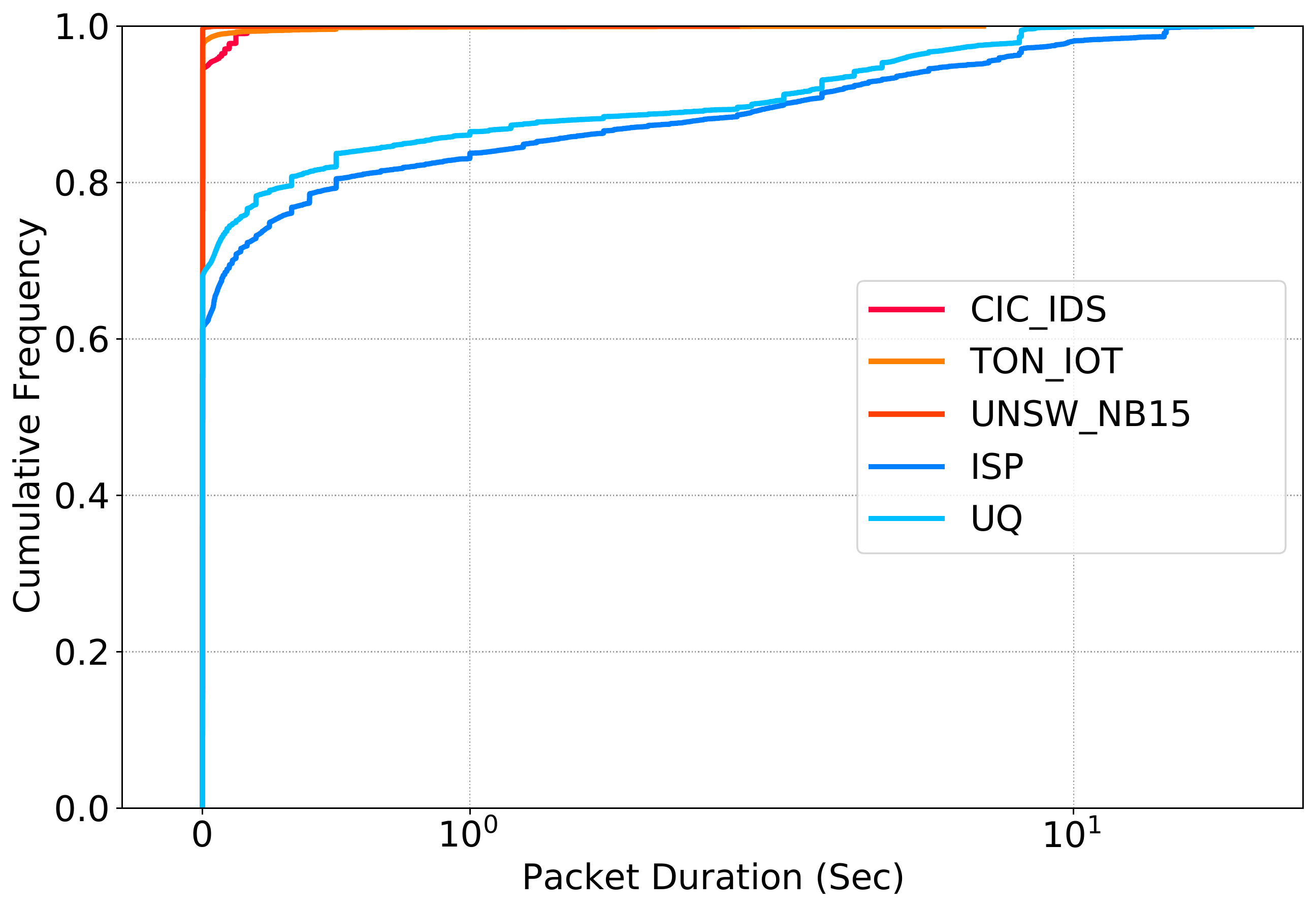}}%
    \caption{Comparing the packet times in five datasets a) boxplots, and b) CDFs}%
    \label{packet time}
\end{figure}

Figure \ref{packet time} shows the distribution of packet times for the five datasets. The average packet time / duration is computed by dividing flow duration by total number of packets in flow, as shown in the Table~\ref{tab: features}. Since it takes into account, at the same time, the flow duration and number of packets, it reflects traffic characteristics in two dimensions time and volume. Again, the similarity between two real-world datasets, and their distance to synthetic datasets is very clear.

\subsubsection{Packet Size}
\begin{figure}[!b]
    \centering
    \subfloat[\centering ]
    {\includegraphics[width=0.45\columnwidth, height=5cm]{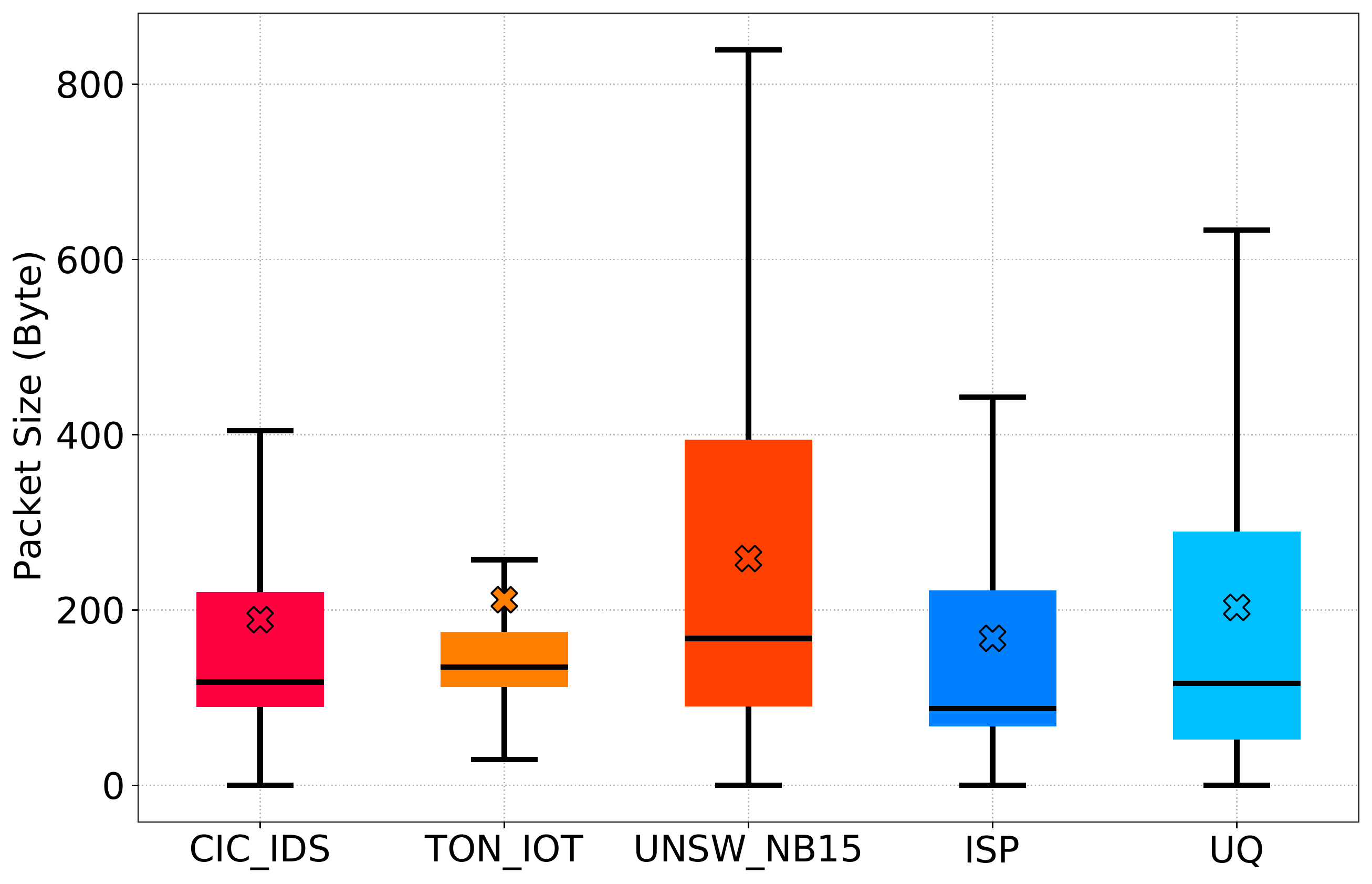} }%
    \qquad
    \subfloat[\centering ]
    {\includegraphics[width=0.45\columnwidth, height=5cm]{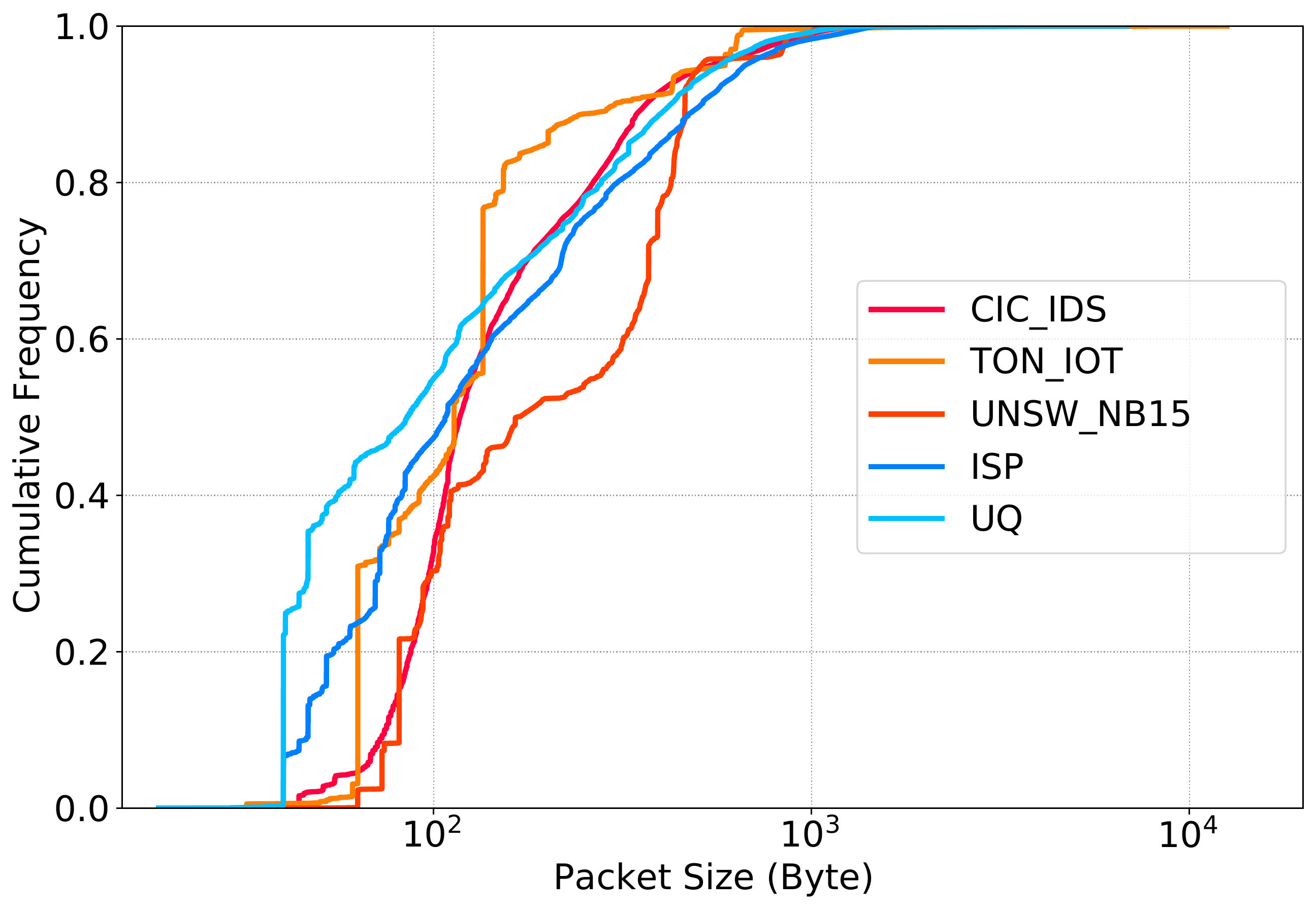}}%
    \caption{Comparing the packet sizes in five datasets a) boxplots, and b) CDFs}%
    \label{packet size}
\end{figure}

Figure \ref{packet size} shows packet size distribution for the five datasets. Although it is hard to separate the real-world and synthetic datasets in Figure \ref{packet size}-a, due to the major overlap between IQRs of both groups, further investigation, as seen in Figure \ref{packet size}-b, reveals a pattern similar to previous features. While the distinction between groups is clear, the difference, in this case, is not as significant as in the case of the previous features. Though, the phenomena of two real-world datasets moving close together, separate from others, from beginning of the CDF till its end, is still observed.
The horizontal axis in Figure~\ref{packet size}-b is again in logarithmic scale, to highlight the main part of the graph, which is compressed due to the outliers with very large values.

\subsubsection{Number of Source IPs per Destination IP}
The next row of the Table \ref{tab: features} indicates a feature that is computed by counting number of source IP addresses per each destination IP address. This feature is specifically important for the detection and identification of DDoS attacks. 
%
%
\begin{figure}[!t]
    \centering
    \subfloat[\centering ]
    {\includegraphics[width=0.45\columnwidth, height=5cm]{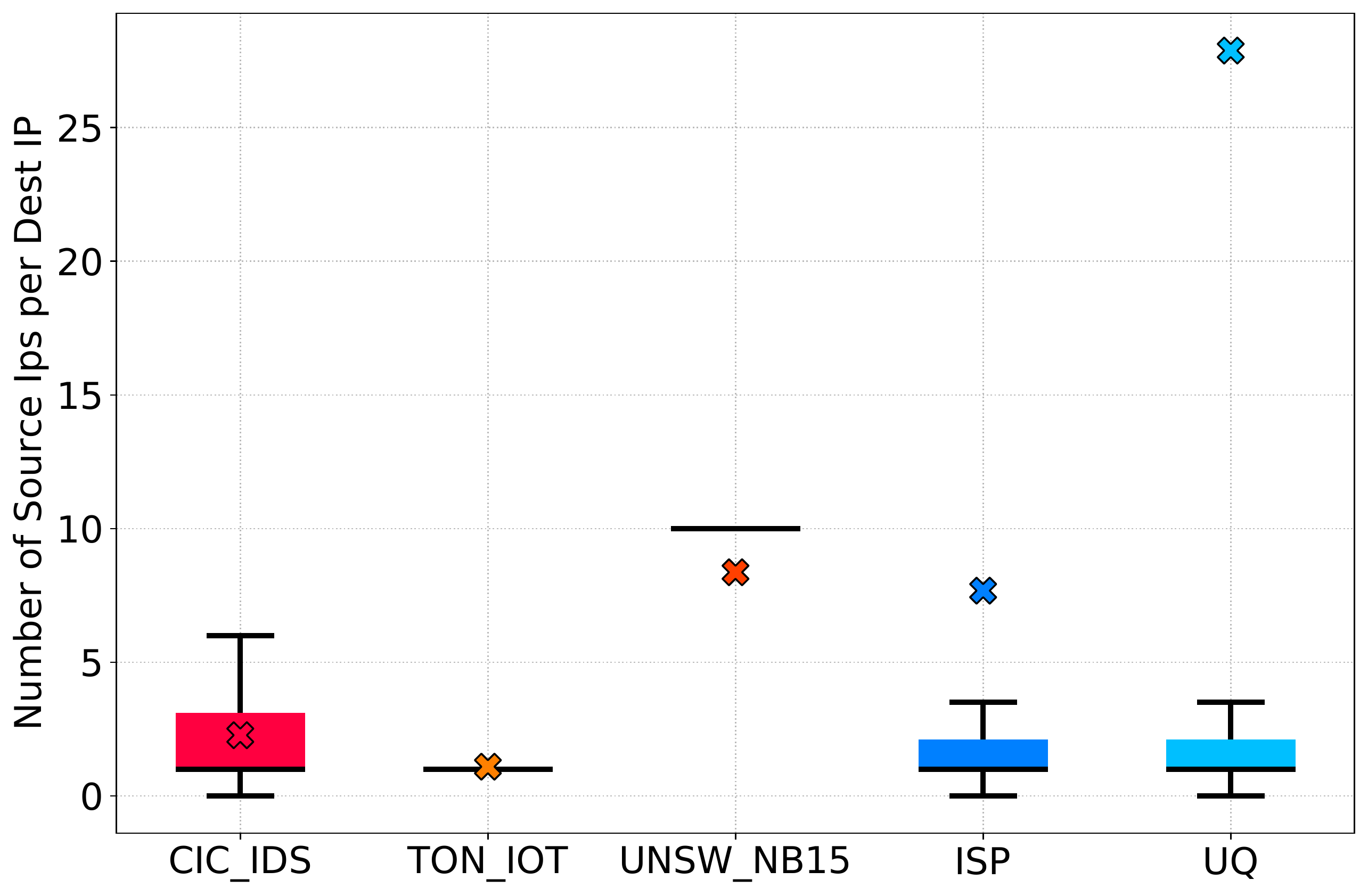} }%
    \qquad
    \subfloat[\centering ]
    {\includegraphics[width=0.45\columnwidth, height=5cm]{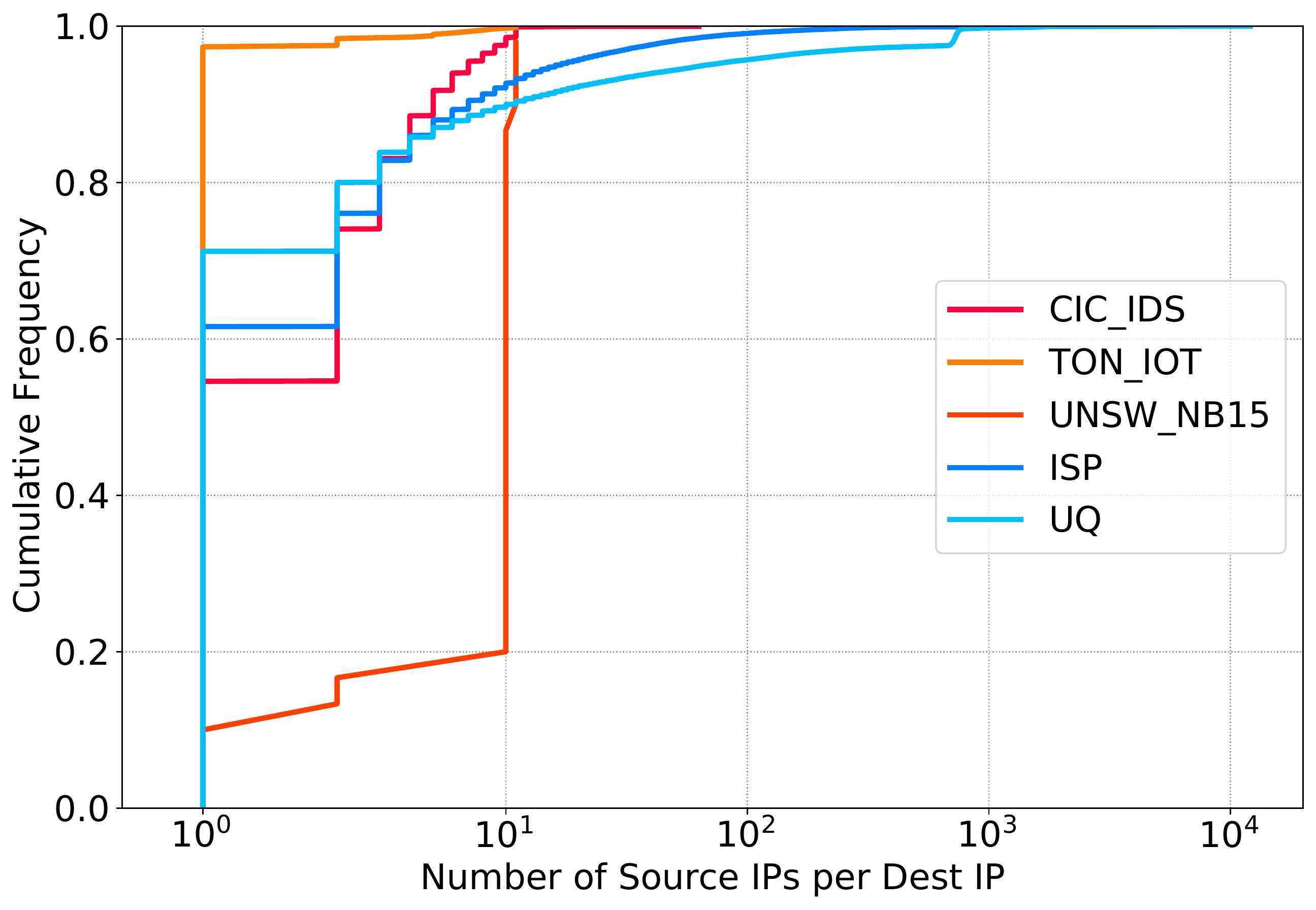}}%
    \caption{Comparing the number of source IPs per destination IP in five datasets a) boxplots, and b) CDFs}%
    \label{num src IP per dest IP}
\end{figure}
%
%
Figure \ref{num src IP per dest IP} shows the distribution of this feature for the five datasets in (a)  using boxplots, and in (b) using CDFs. As in Figure \ref{num src IP per dest IP}-a, the two real-world datasets have a completely similar IQR that is totally different from UNSW\_NB15 and TON\_IOT, and the IQR of CIC dataset is overlapping with the real-world datasets. Figure \ref{num src IP per dest IP}-b also confirms this by showing that the CDF curves of the two real-world datasets move close together from beginning to the end. The CDF curve of CIC starts close to the real-world datasets but somewhere in the end of the range separates from them. However, the CDF curves of the UNSW\_NB15 and TON\_IOT take a distinct path from early beginning to the end. These results are completely inline with the previous results, i.e. results of the previous features. As mentioned earlier, the results in this section includes the initial analysis and we have done the precise statistical tests for all feature distribution comparisons, that their results will be provided the results in the next subsection.

\begin{figure}[!b]
    \centering
    \subfloat[\centering ]
    {\includegraphics[width=0.45\columnwidth, height=5cm]{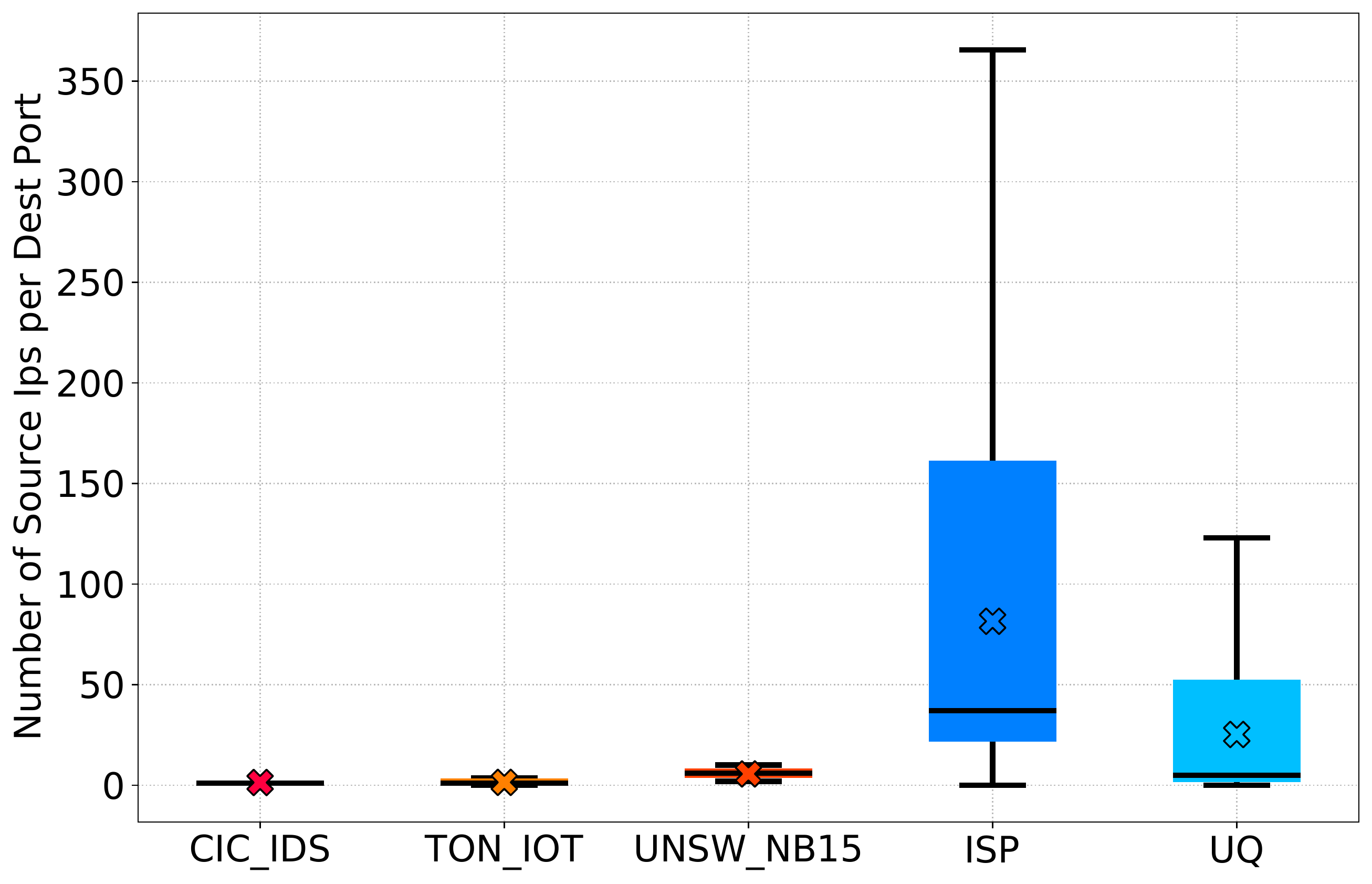} }%
    \qquad
    \subfloat[\centering ]
    {\includegraphics[width=0.45\columnwidth, height=5cm]{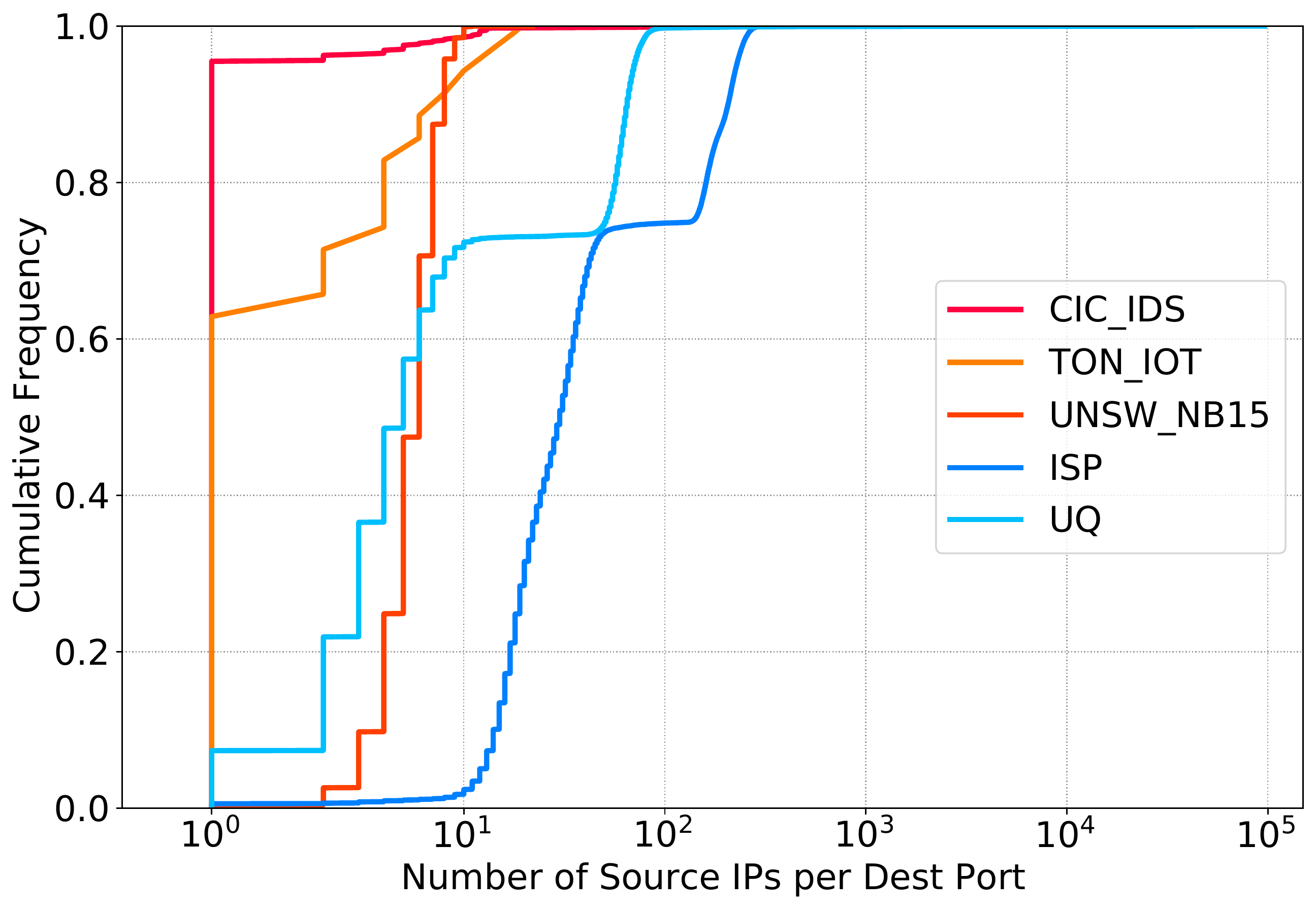}}%
    \caption{Comparing the number of source IP addresses per destination port in five datasets a) boxplots, and b) CDFs}%
    \label{num src IP per dest Port}
\end{figure}

\subsubsection{Number of Source IPs per Destination Port}
This feature is again important in detecting DoS and DDoS and scanning attacks. The feature distribution for the five datasets are depicted in Figure \ref{num src IP per dest Port}-a in boxplots and in Figure \ref{num src IP per dest Port}-b in CDFs. The distinction between the real-world and synthetic datasets is very clear in both boxplots and CDF curves. In Figure \ref{num src IP per dest Port}-a, the three synthetic datasets (red boxplots) have very small IQR, in contrast to two real-world datasets (blue boxplots), which have large overlapped IQR. The CDF curves have exactly the same situation, i.e. the two real-world datasets are relatively close together from beginning  all the way to the end, and distinct from synthetic datasets.

\begin{figure}[!b]
    \centering
    \subfloat[\centering ]
    {\includegraphics[width=0.45\columnwidth, height=5cm]{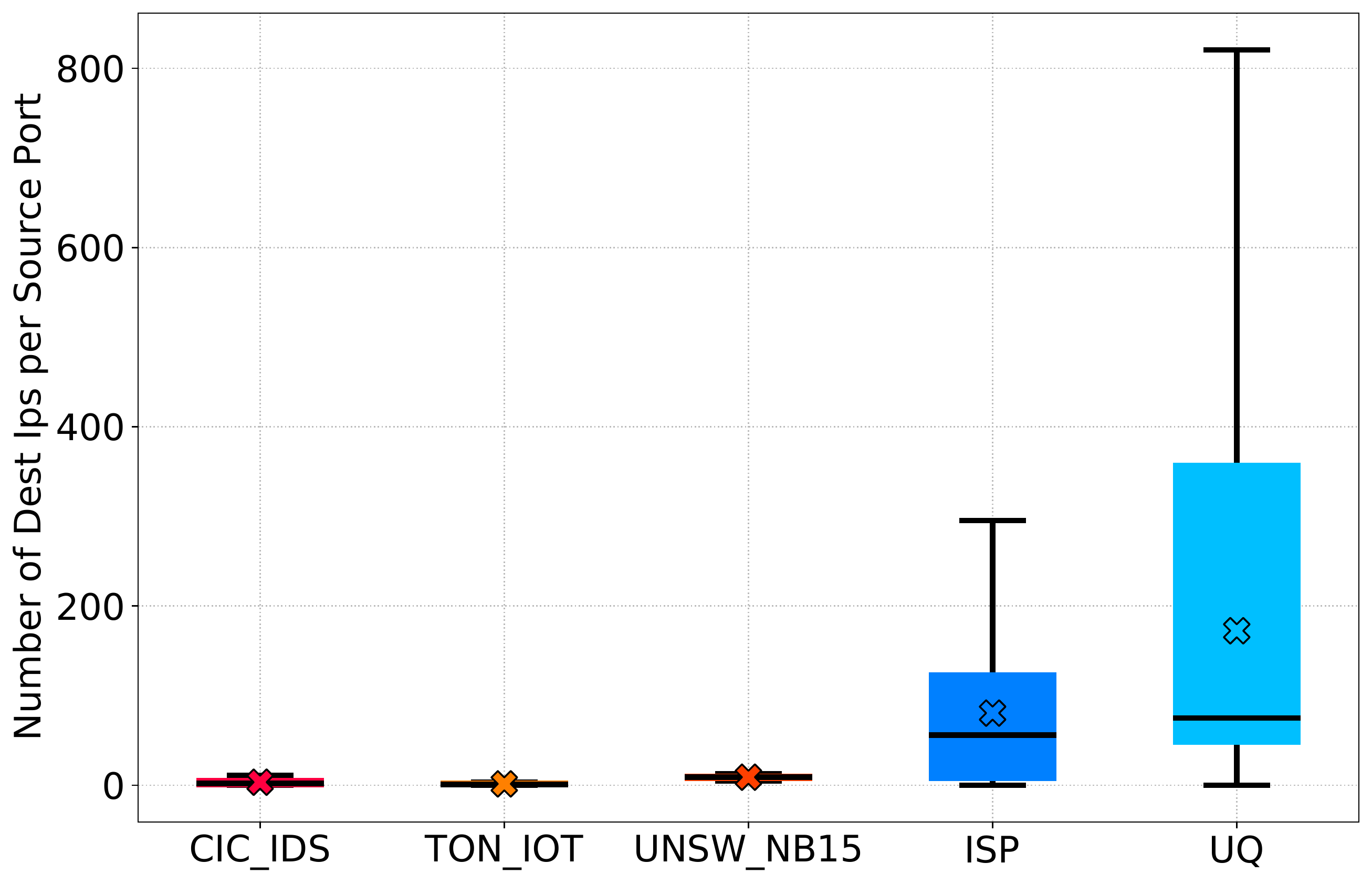} }%
    \qquad
    \subfloat[\centering ]
    {\includegraphics[width=0.45\columnwidth, height=5cm]{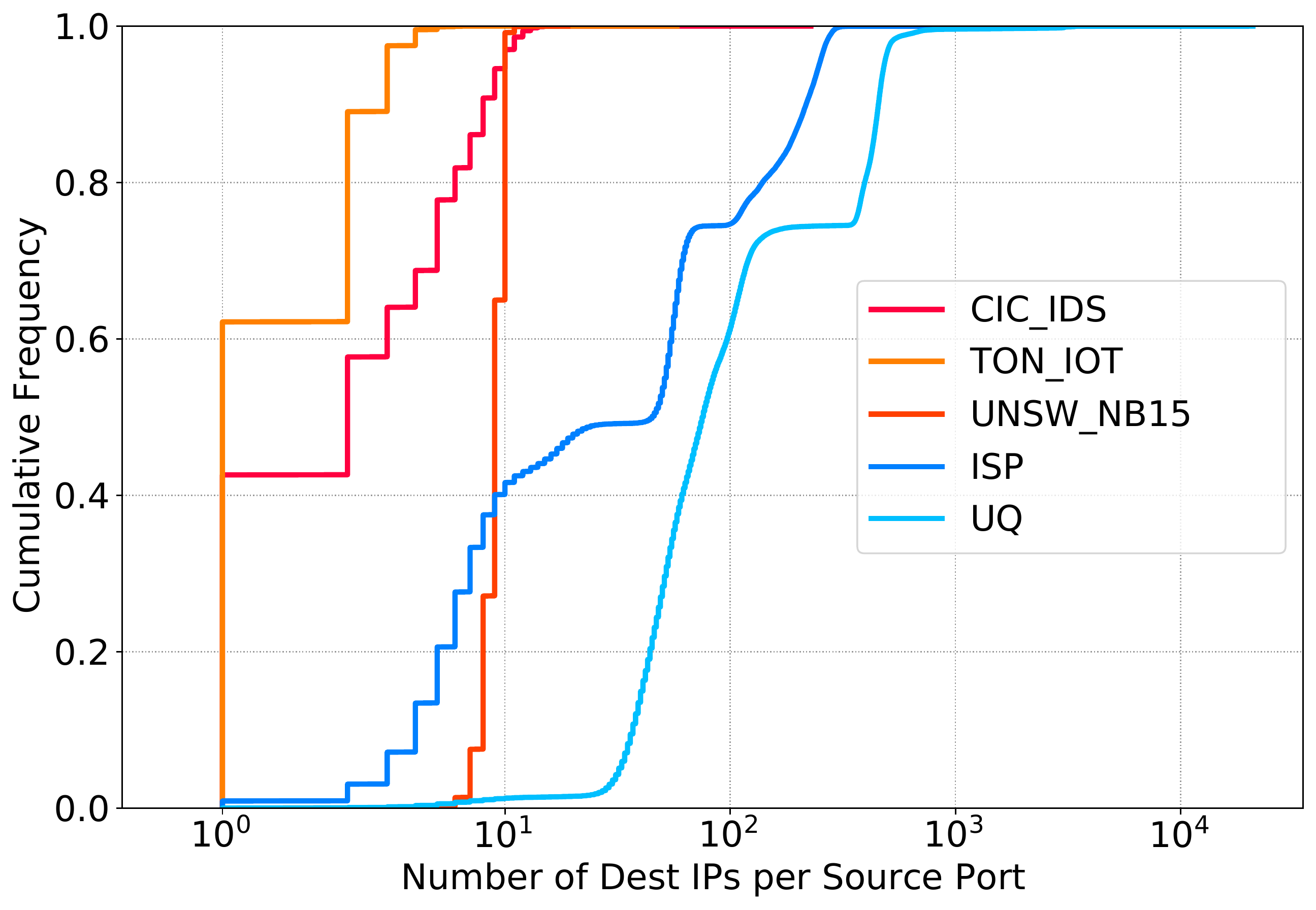}}%
    \caption{Comparing the number of destination IP addresses per source port in five datasets a) boxplots, and b) CDFs}%
    \label{num dest IP per src Port}
\end{figure}

\subsubsection{Number of Destination IPs per Source Port}


Again, this feature is important in detecting DoS, DDoS and scanning attacks. The boxplots and CDF curves of the feature distribution for the five datasets are depicted in Figure \ref{num dest IP per src Port}-a and \ref{num dest IP per src Port}-b respectively. 
The situation of feature distribution is completely similar to the previous feature. The two real-world datasets have very close distributions in both boxplots and CDFs, and they are distinct from the synthetic datasets.

\subsubsection{Number of Destination Ports per Source Port}
\begin{figure}[!t]
    \centering
    \subfloat[\centering ]
    {\includegraphics[width=0.45\columnwidth, height=5cm]{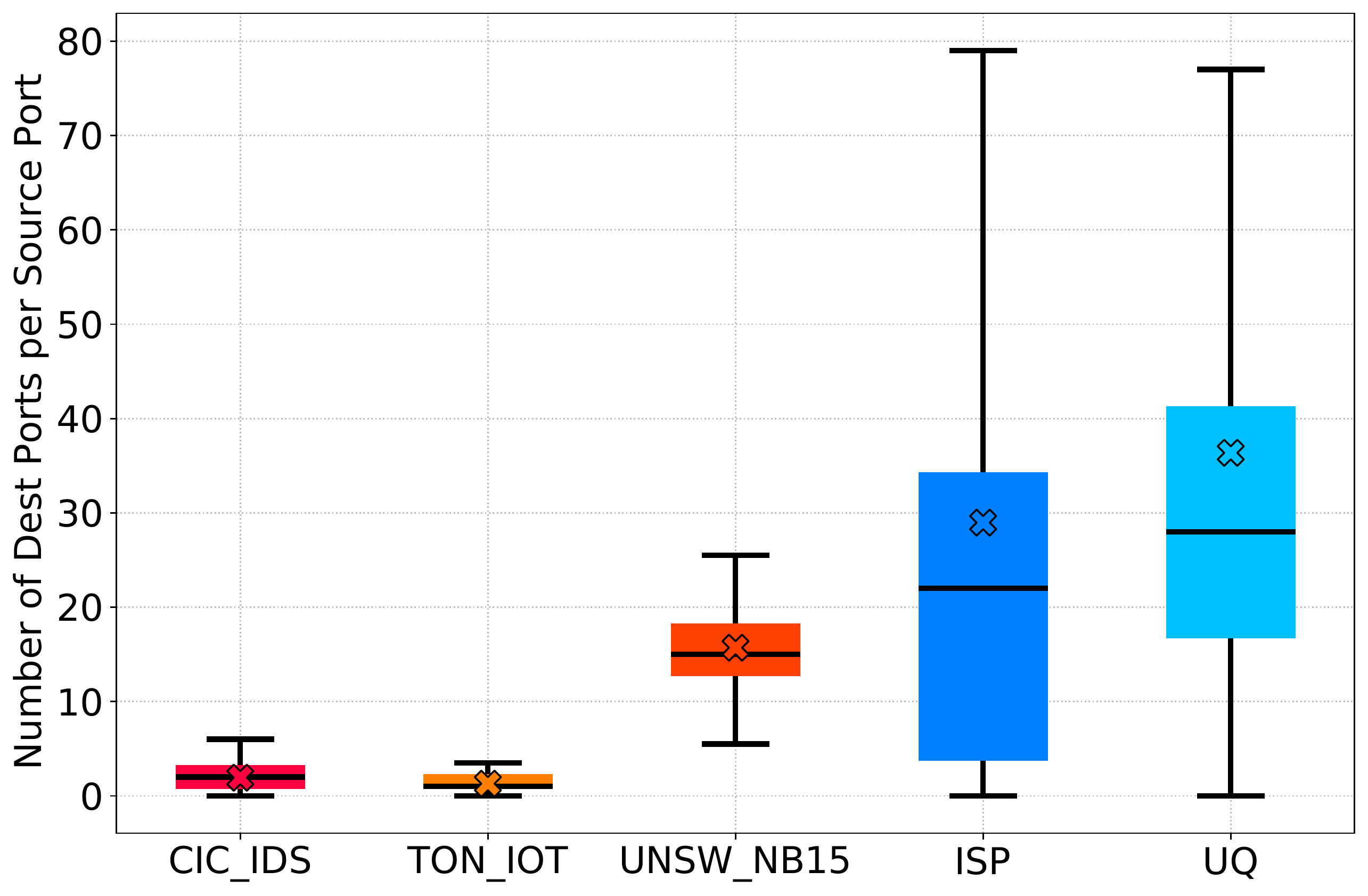} }%
    \qquad
    \subfloat[\centering ]
    {\includegraphics[width=0.45\columnwidth, height=5cm]{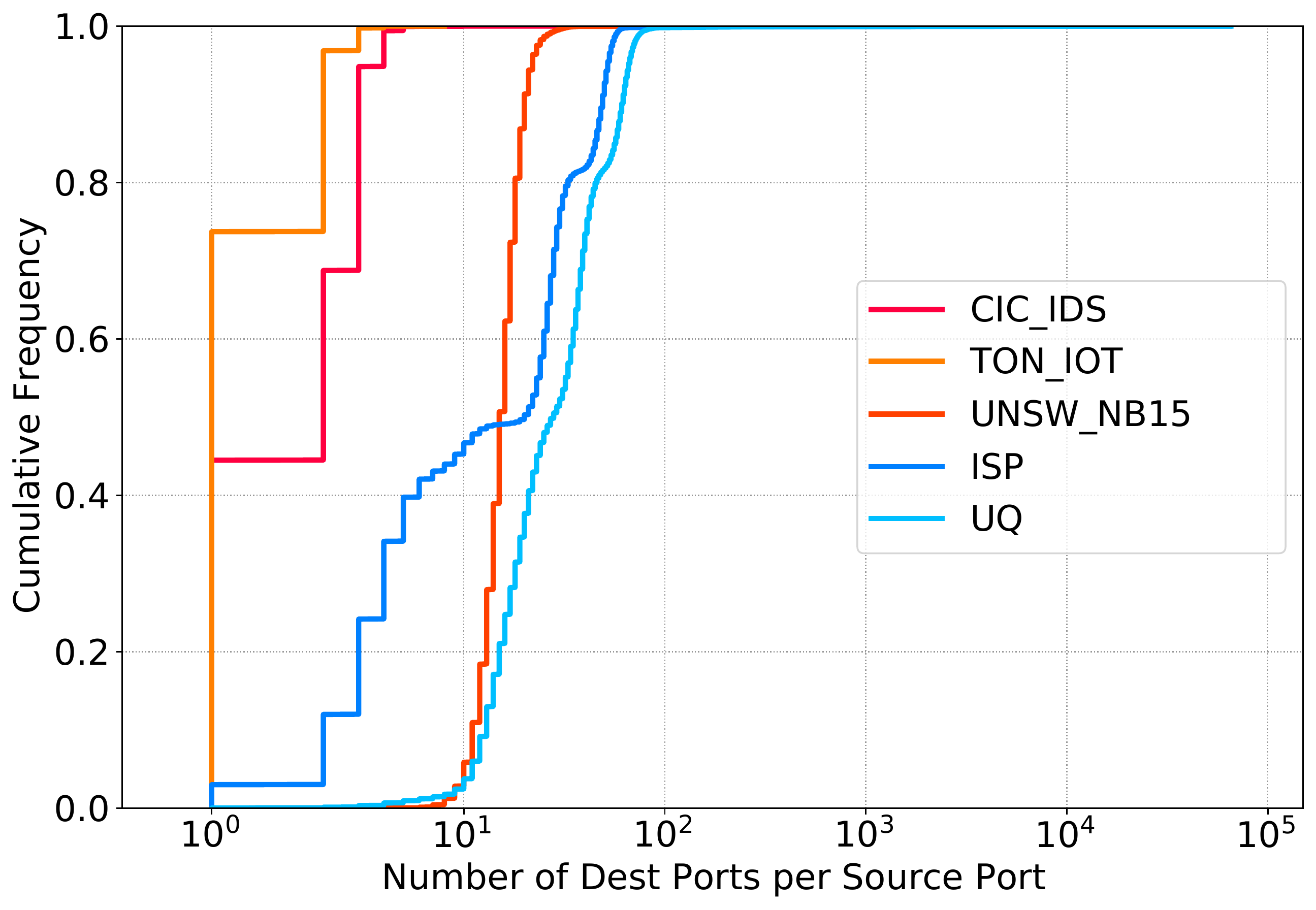}}%
    \caption{Comparing the number of destination ports per source port in five datasets a) boxplots, and b) CDFs}%
    \label{num dest Port per src Port}
\end{figure}

This feature is also important in detecting DDoS and scanning attacks. The feature distribution for the five datasets are depicted in boxplots (Figure \ref{num dest Port per src Port}-a ) and CDF curves (Figure \ref{num dest Port per src Port}-b). 
Like the two previous features, the distributions of the feature in two real-world datasets are very close in both boxplots and CDFs, and they are distinct from the synthetic datasets with an exception of UNSW\_NB15 which gets close to the real-world datasets in parts of its CDF curve.

\subsubsection{Number of L7 Protocols per Destination Port}
This is the last feature used in this study for the comparison of the synthetic datasets with the real-world traffic. While due to the lack of availability of L7 protocol in the fields exported by NetFlow exporters in the past, it has not been utilised in the anomaly detection studies referred in our study, it is an important feature in network security. Intuitively, many well-known protocols have been used with specific destination ports, such as the HTTP and port 80. As such, the number of L7 protocols utilised with each port can be attributed to the normal behavior, i.e. characteristics, of a network and changes in its distribution can be an indicator of abnormal network situation.

\begin{figure}[!b]
    \centering
    \subfloat[\centering ]
    {\includegraphics[width=0.45\columnwidth, height=5cm]{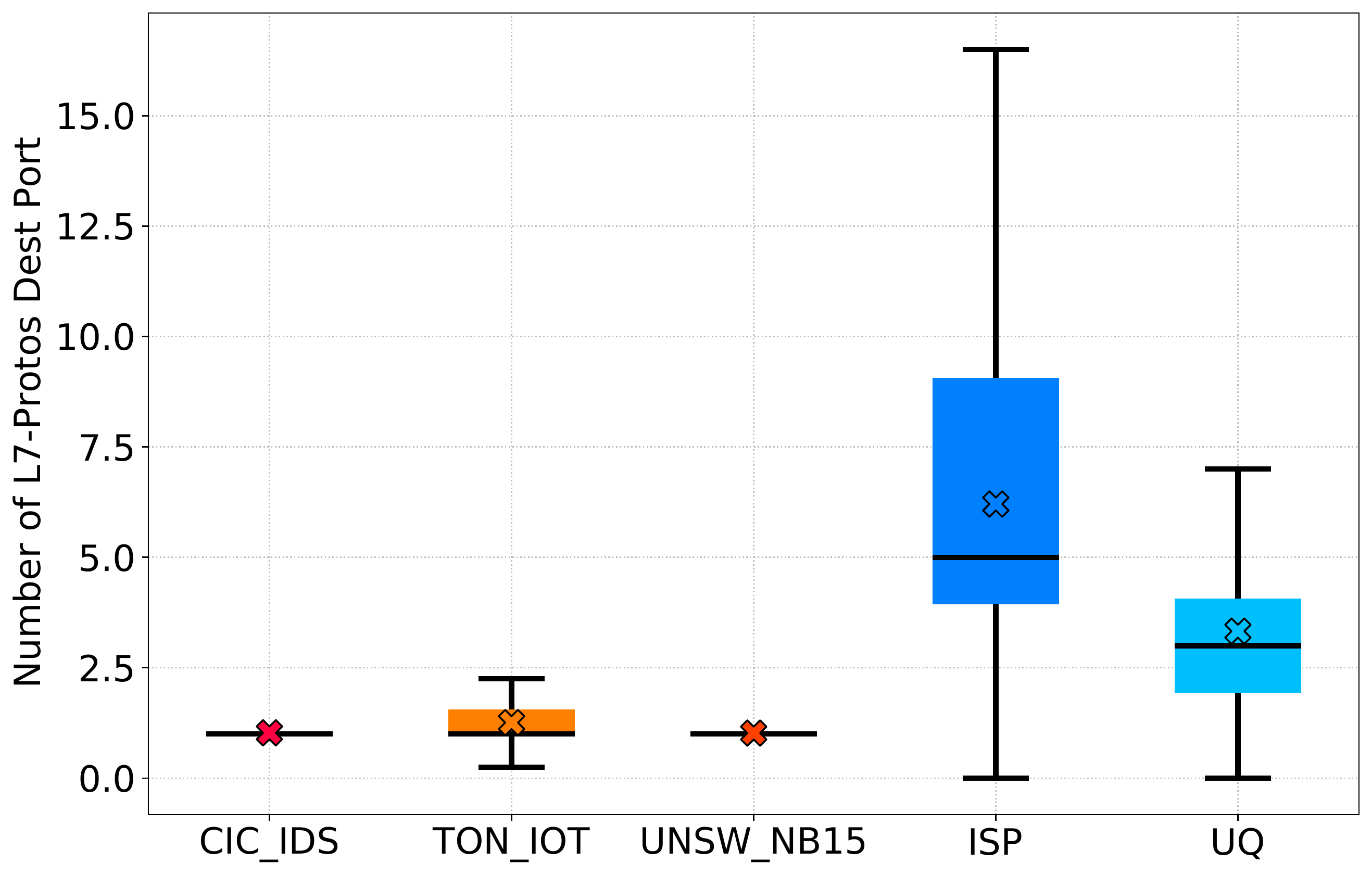} }%
    \qquad
    \subfloat[\centering ]
    {\includegraphics[width=0.45\columnwidth, height=5cm]{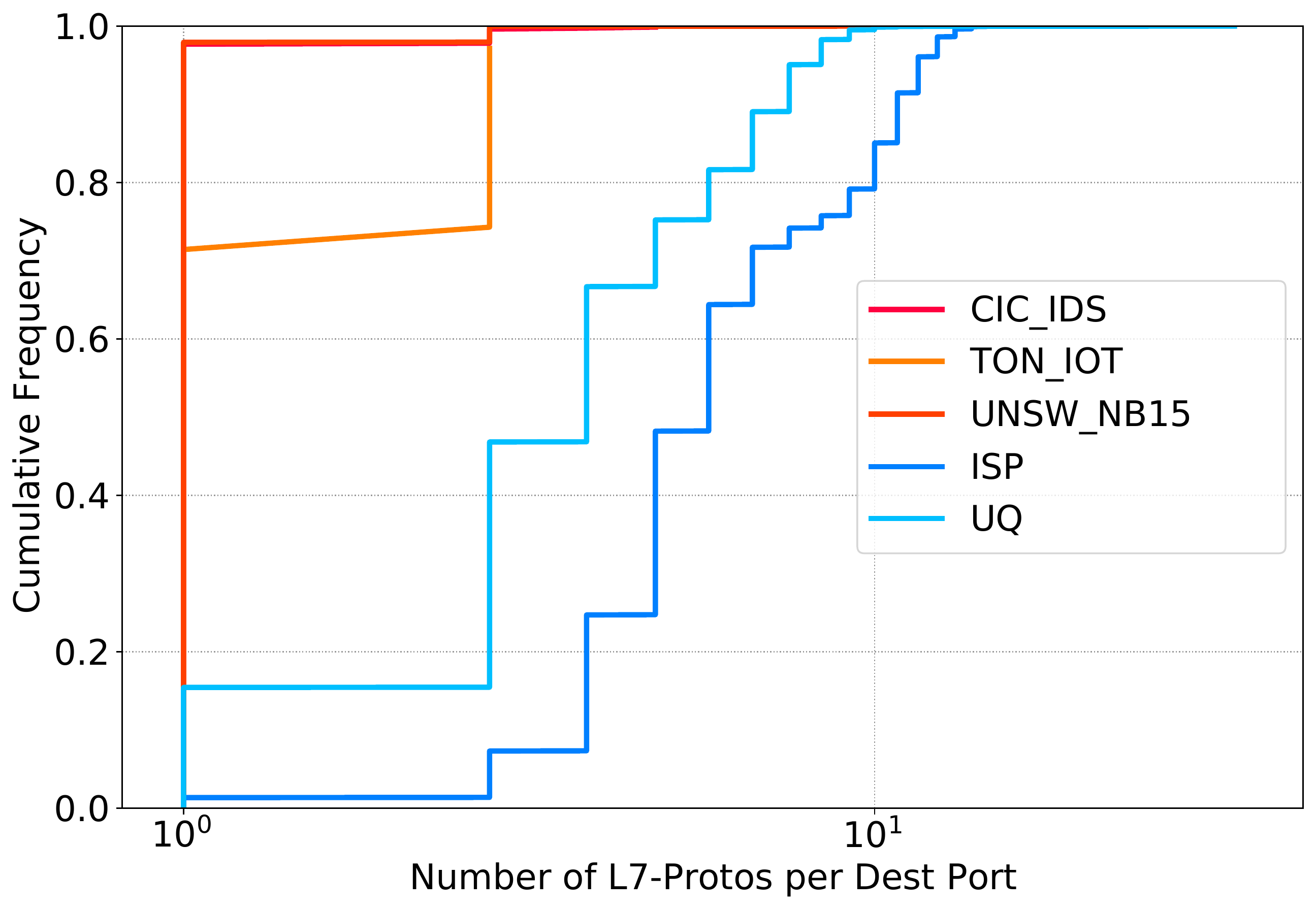}}%
    \caption{Comparing the number of L7 protocols per destination port in five datasets a) boxplots, and b) CDFs}%
    \label{num L7 per dst Port}
\end{figure}

The distribution of this feature for the five datasets are shown in Figure \ref{num L7 per dst Port}. The boxplots are shown in Figure \ref{num L7 per dst Port}-a in which the three synthetic datasets are depicted in red color in the left most side and the two real-world datasets are depicted in blue in the right-most side of the figure. As seen, the IQRs of the two real-world datasets are totally separate from the synthetic datasets. Although the real-world datasets seem to have non-overlapping IQRs, their both differences to synthetic datasets' IQRs are stronger than their difference. Similar situation can be understood from the CDF curves as shown in Figure \ref{num L7 per dst Port}-b that confirms relative closeness of the real-world datasets compared to synthetic datasets. While this intuitive analysis gives us a sense of the feature distribution comparison between these five datasets, the results provided by running Kruskal-Wallis test in the next subsection will provide the final findings.

\subsection{Comparing Dimensionality Reduced Feature Distributions} 
\begin{figure}[!t]
    \centering
    \subfloat[\centering ]
    {\includegraphics[width=0.45\columnwidth, height=5cm]{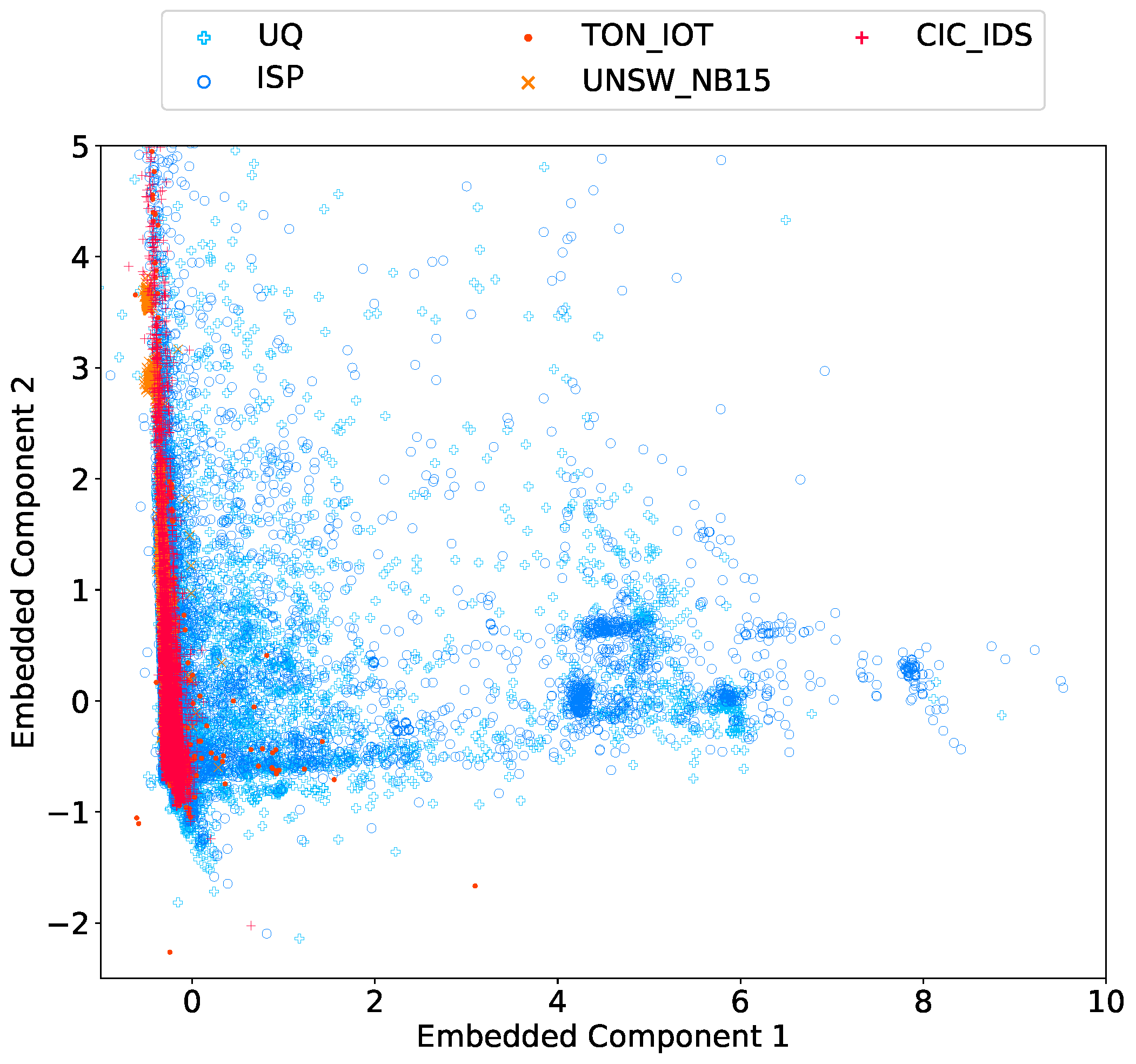}}%
    \hspace{0.5cm}
    \subfloat[\centering ]
    {\includegraphics[width=0.45\columnwidth, height=5cm]{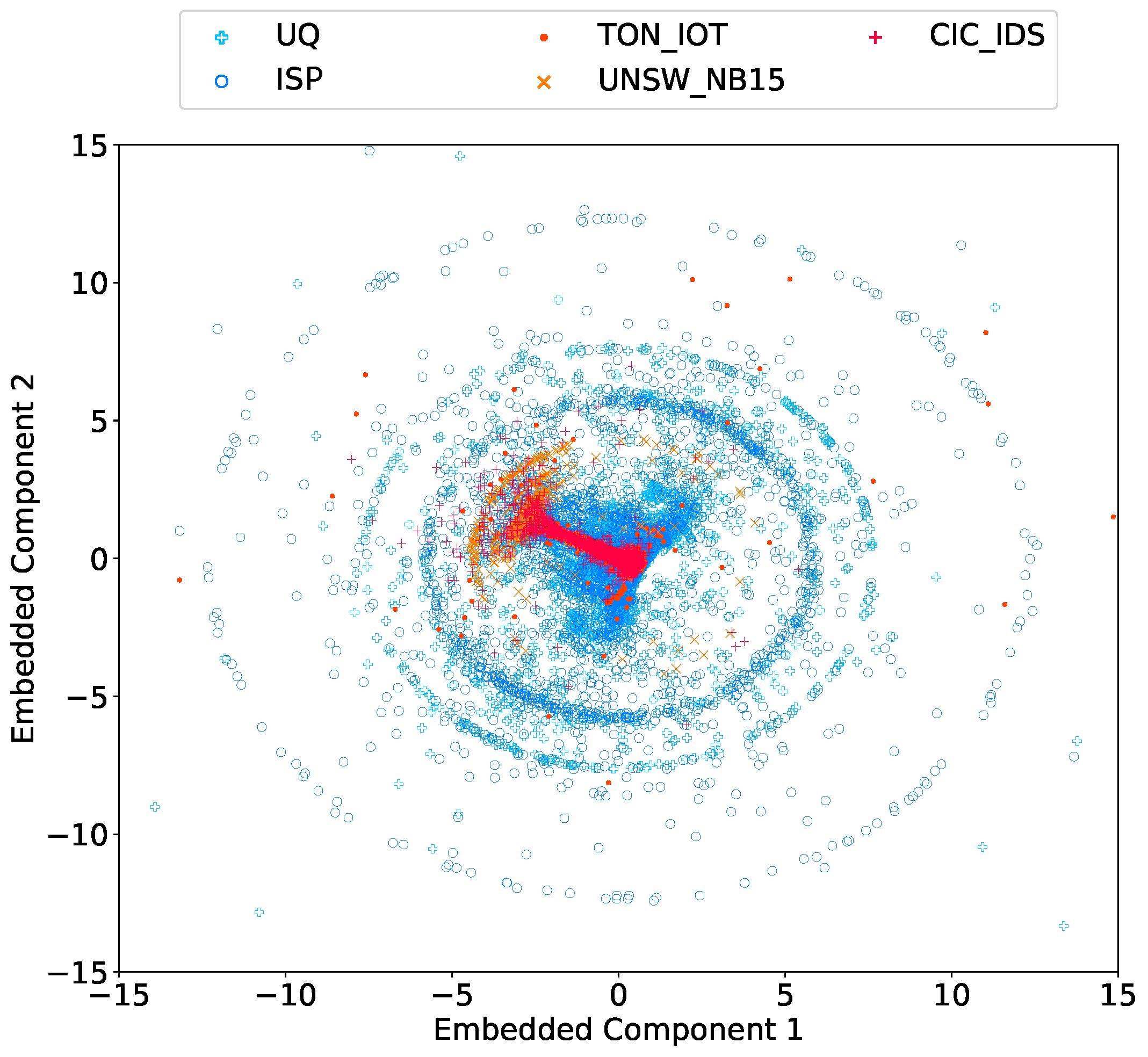}}%
    \qquad
        \subfloat[\centering ]
    {\includegraphics[width=0.45\columnwidth, height=5cm]{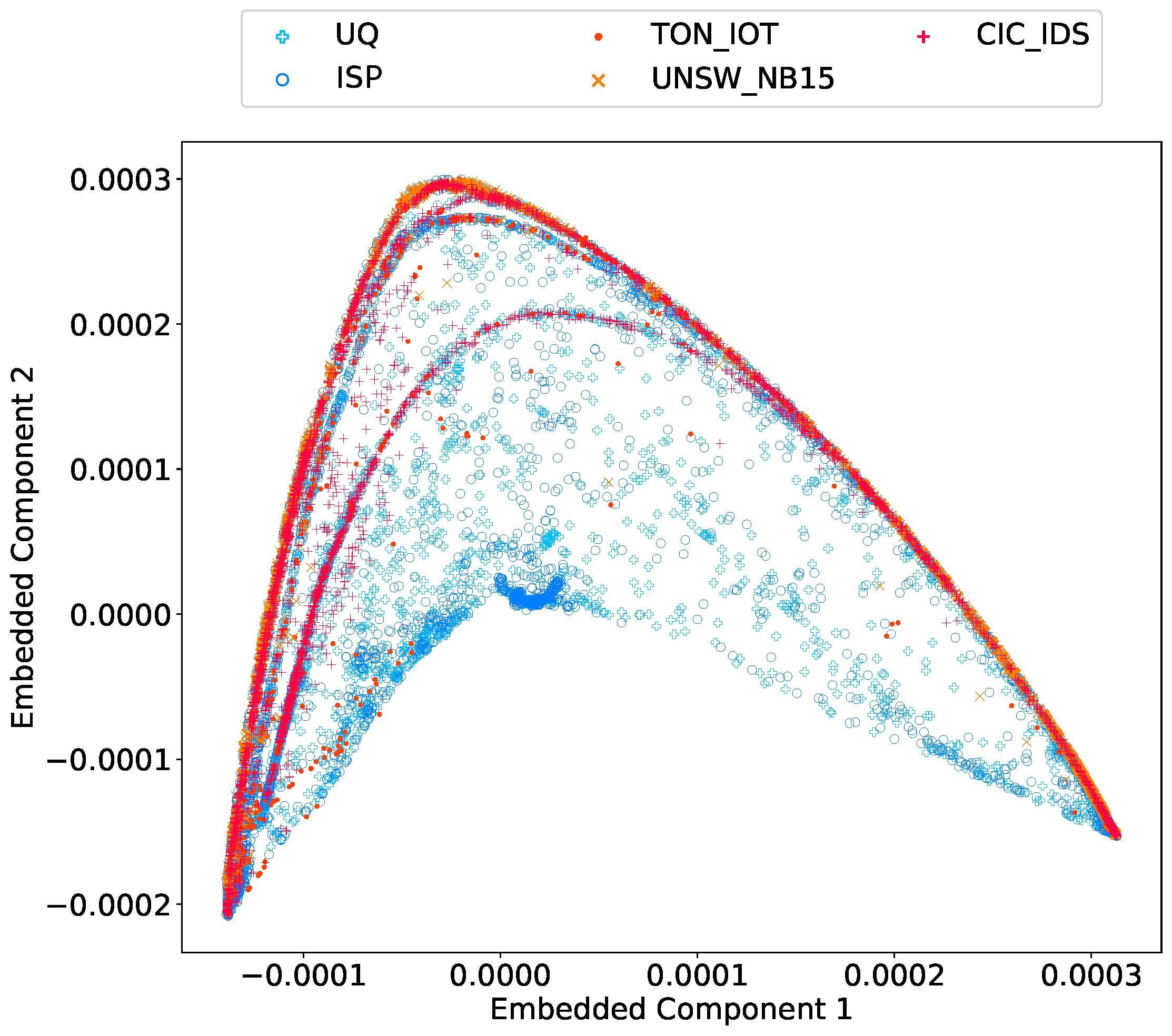}}%
    \hspace{0.5cm}
        \subfloat[\centering ]
    {\includegraphics[width=0.45\columnwidth,
    height=5cm]{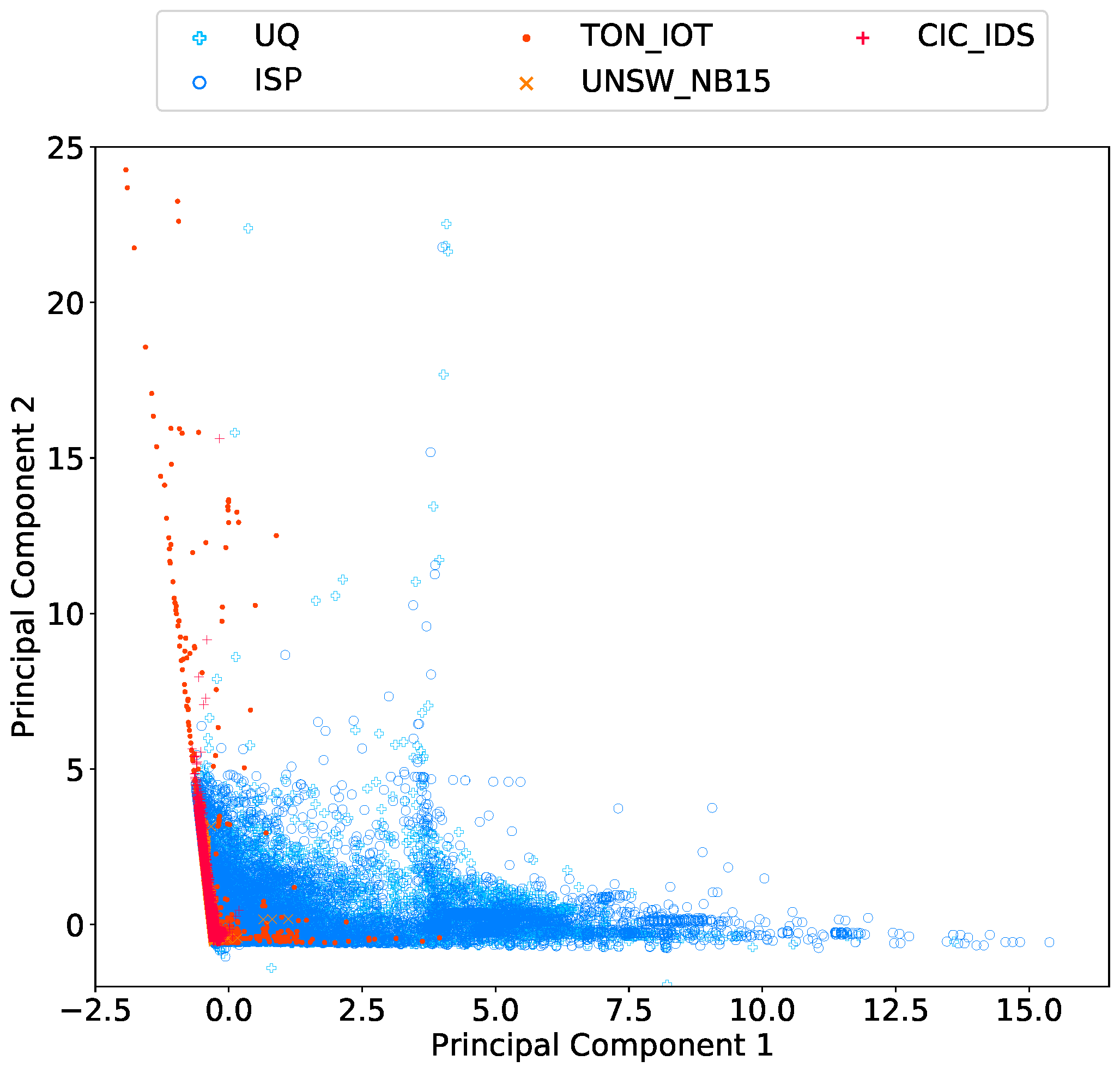}}%

    \caption{Embedding samples all the five datasets using a) Linear Discriminant Analysis (LDA), b) Multi-dimensional Scaling(MDS), c) Spectral Embedding, and d) Principle Component Analysis (PCA)}%
    \label{embeddings}
\end{figure}

Here we visualize the feature distributions after applying various methods of embedding, on the set of all nine features listed in Table~\ref{tab: features},   and compare the five datasets. For this purpose, we use four different methods of dimensionality reduction techniques including Linear Discriminant Analysis (LDA), Multi-dimensional Scaling(MDS), Spectral Embedding and Principle Component Analysis (PCA). The resulting embeddings of features after applying dimensionality reduction on each dataset are plotted in Figure~\ref{embeddings}-a, b, c and d respectively. Embeddings of each dataset is plotted using a different marker and color to illustrate the differences of their distributions.

In Figure~\ref{embeddings}-a, LDA is the dimensionality reduction method applied to the features. The horizontal axis represents the first embedded component and the vertical axis represents the second embedded component. As per previous figures, dark and light blue colors are used for UQ and ISP and the orange, light and dark red are used for UNSW_NB15, TON_IOT and CIC_IDS respectively. As seen, the embedded points from the two real-world datasets are vastly spread in the direction of the first embedded component, while the embedded points of the synthetic datasets are mostly located in the area of small values. Similar results can be seen in  Figure~\ref{embeddings}-d, where PCA is applied for dimensionality reduction and the horizontal and vertical axes represent the first and second principle components respectively.

The other two embedding show another pattern in which the real-world and synthetic datasets are separated across both embedded components. In Figure~\ref{embeddings}-b and Figure~\ref{embeddings}-c, which illustrate the results of MDS and spectral embedding  dimensionality reduction methods, the horizontal and vertical axes represent the first and second embedded component respectively. In both figures, the embedded points of the synthetic datasets are mostly accumulated in a small area of the surface, while the embedded points of the real-world datasets are spread in much larger area. These results are another indication of the vast differences between the statistical features of the real-world datasets and the \textbf{normal} part of these synthetic datasets.
\\
\\

\section{Quantifying the Distance Between Distributions}\label{Analysis}
In the previous section we have shown that distributions of  traffic characteristics features of the benign / attack-free parts of the three selected IDS datasets are statistically different from the real-world traffic. This distribution difference is visualised for several features that are commonly used for implementing anomaly and intrusion detection systems, in terms of their CDF and boxplot comparison. 
While these visualizations are clear indications of the difference between the two groups of datasets, they do not provide a measure of how an individual dataset is far from the other one. 

\begin{figure}[!t]
    \centering
    \subfloat[\centering ]
    {\includegraphics[width=5cm]
    {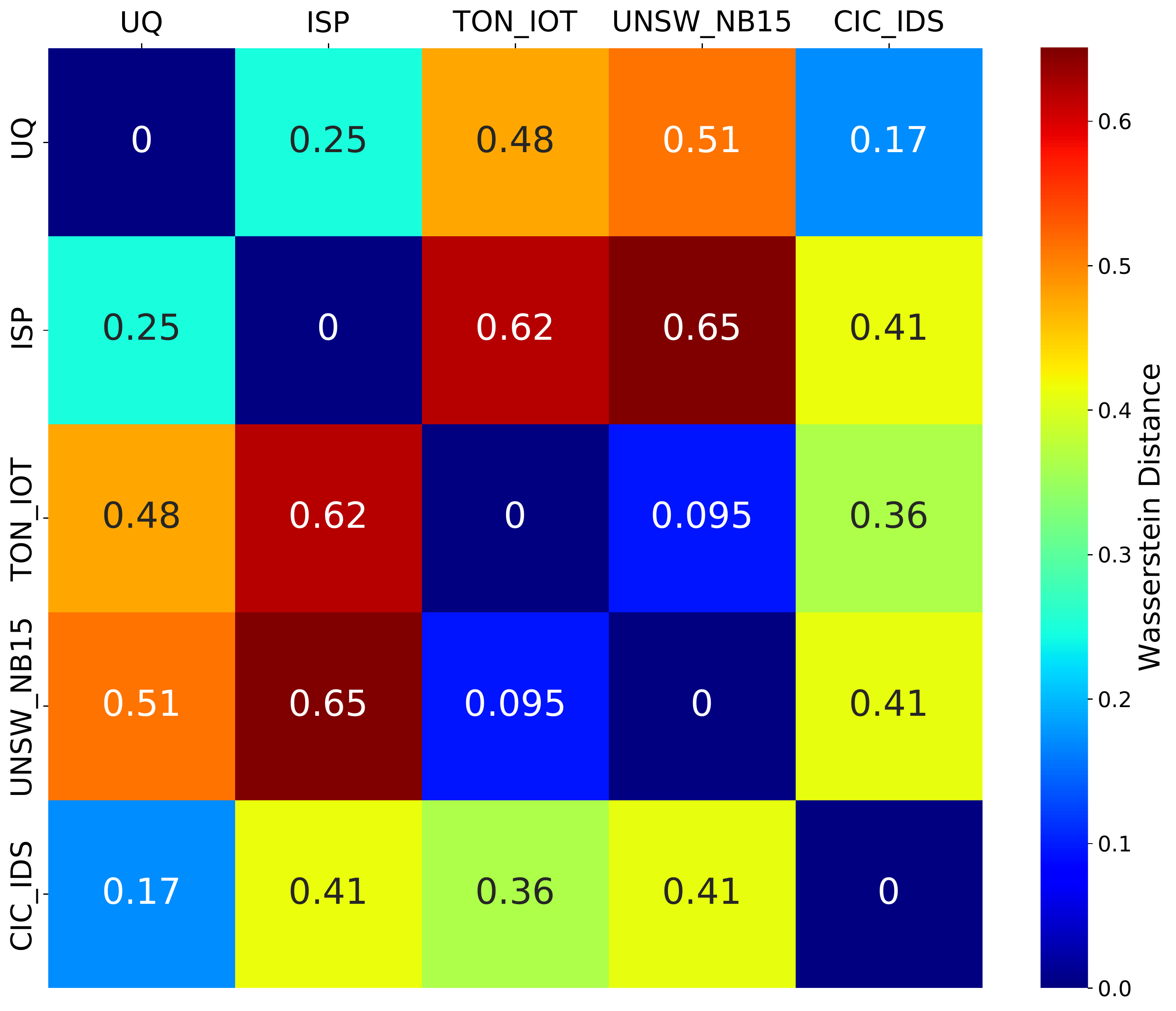}}%
     \hspace{0.25cm}
    \subfloat[\centering ]
    {\includegraphics[width=5cm]
    {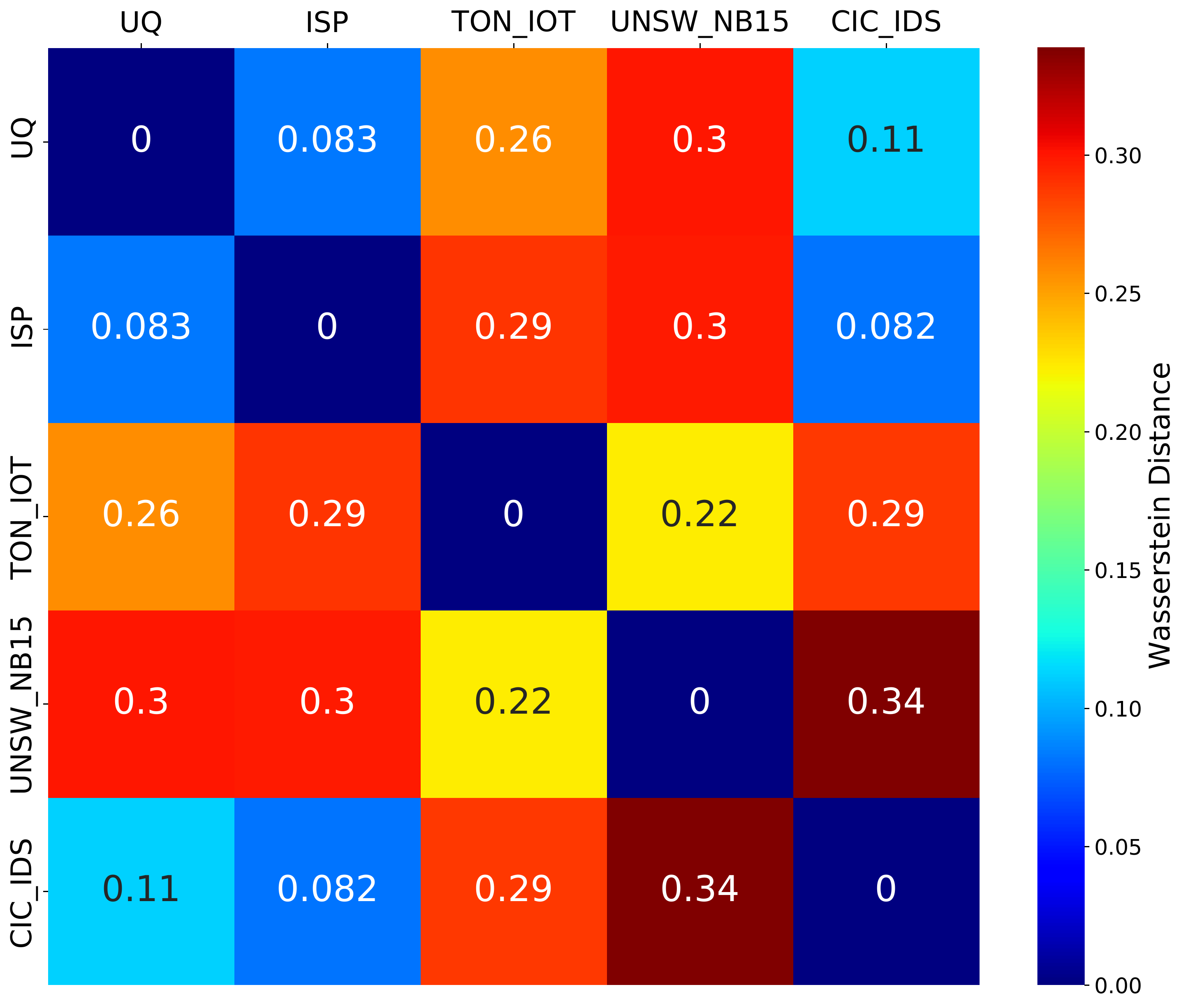}}%
    \hspace{0.25cm}         
    \subfloat[\centering ]
    {\includegraphics[width=5cm]
    {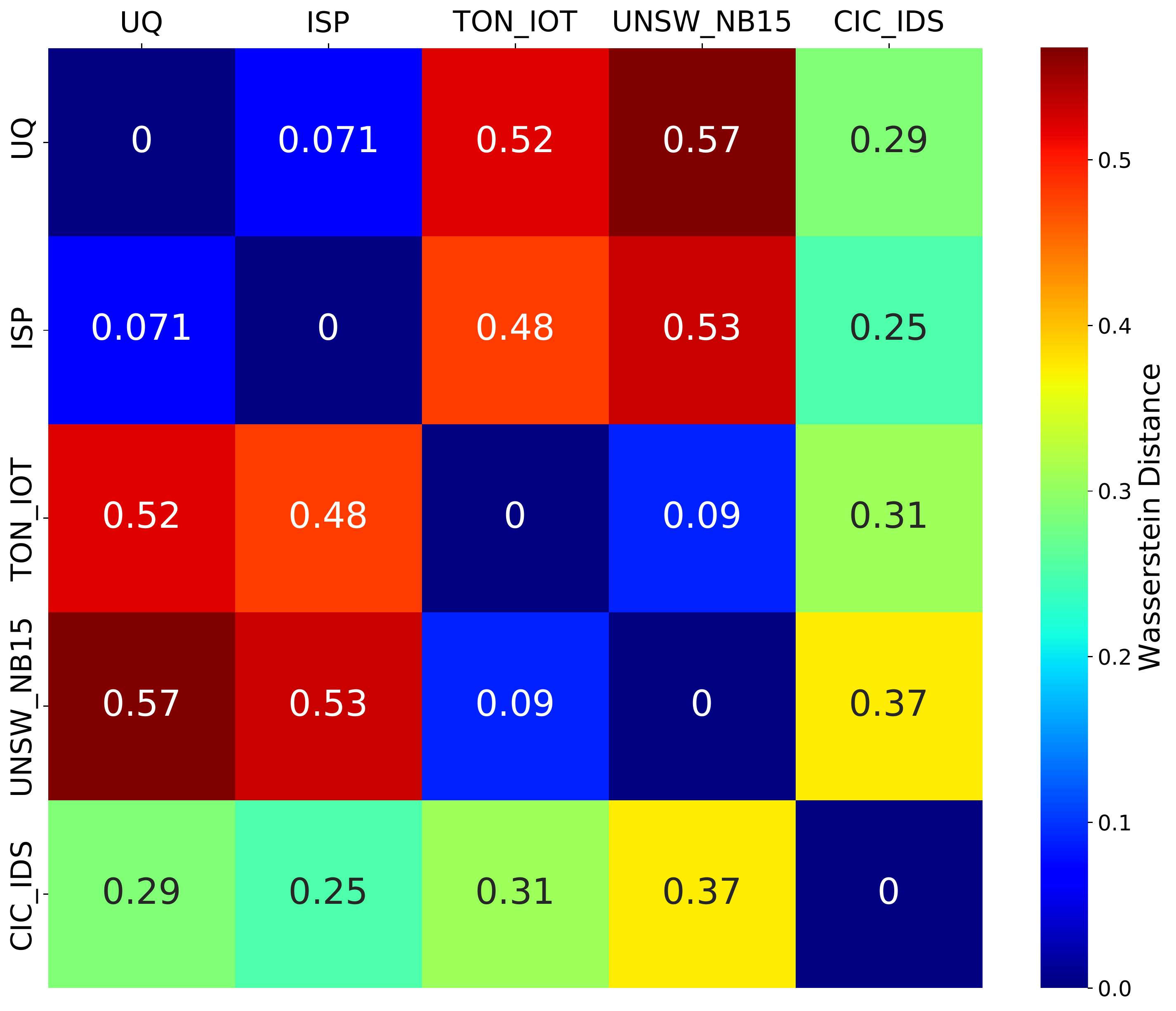}}%
     \qquad
    \subfloat[\centering ]
    {\includegraphics[width=5cm]
    {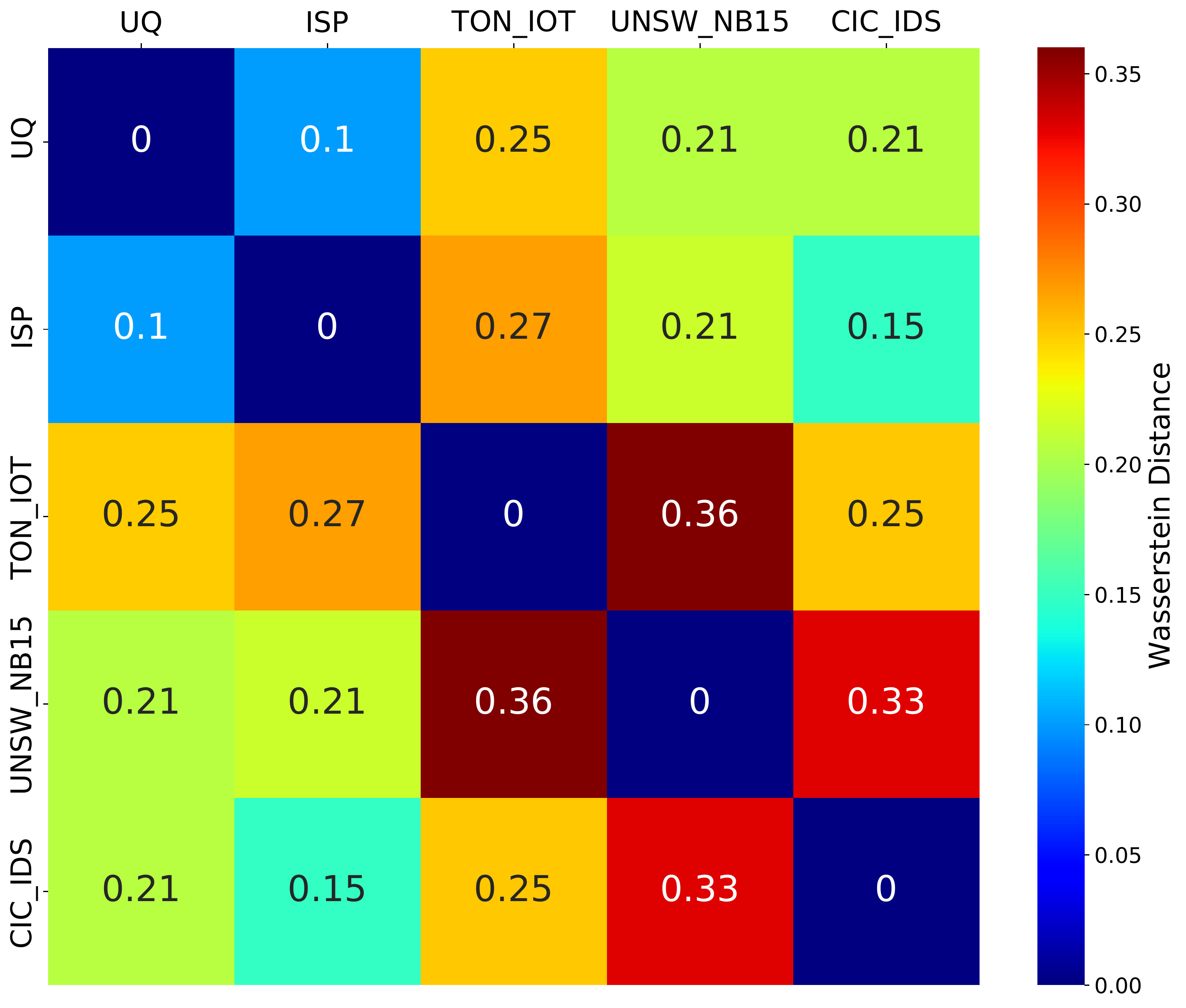}}%
    \hspace{0.25cm}
    \subfloat[\centering ]
    {\includegraphics[width=5cm]
    {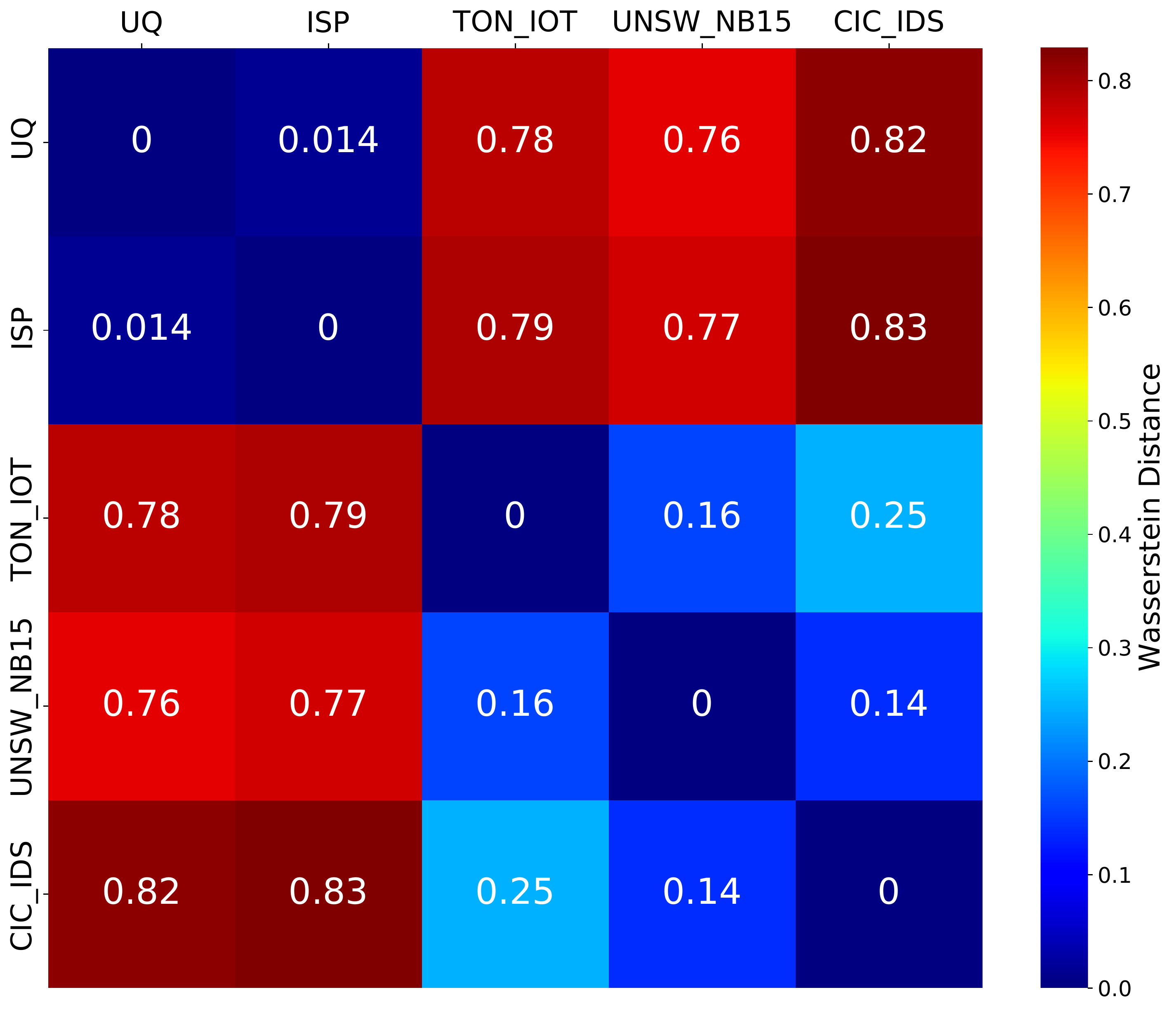}}%
    \hspace{0.25cm}
    \subfloat[\centering ]
    {\includegraphics[width=5cm]
    {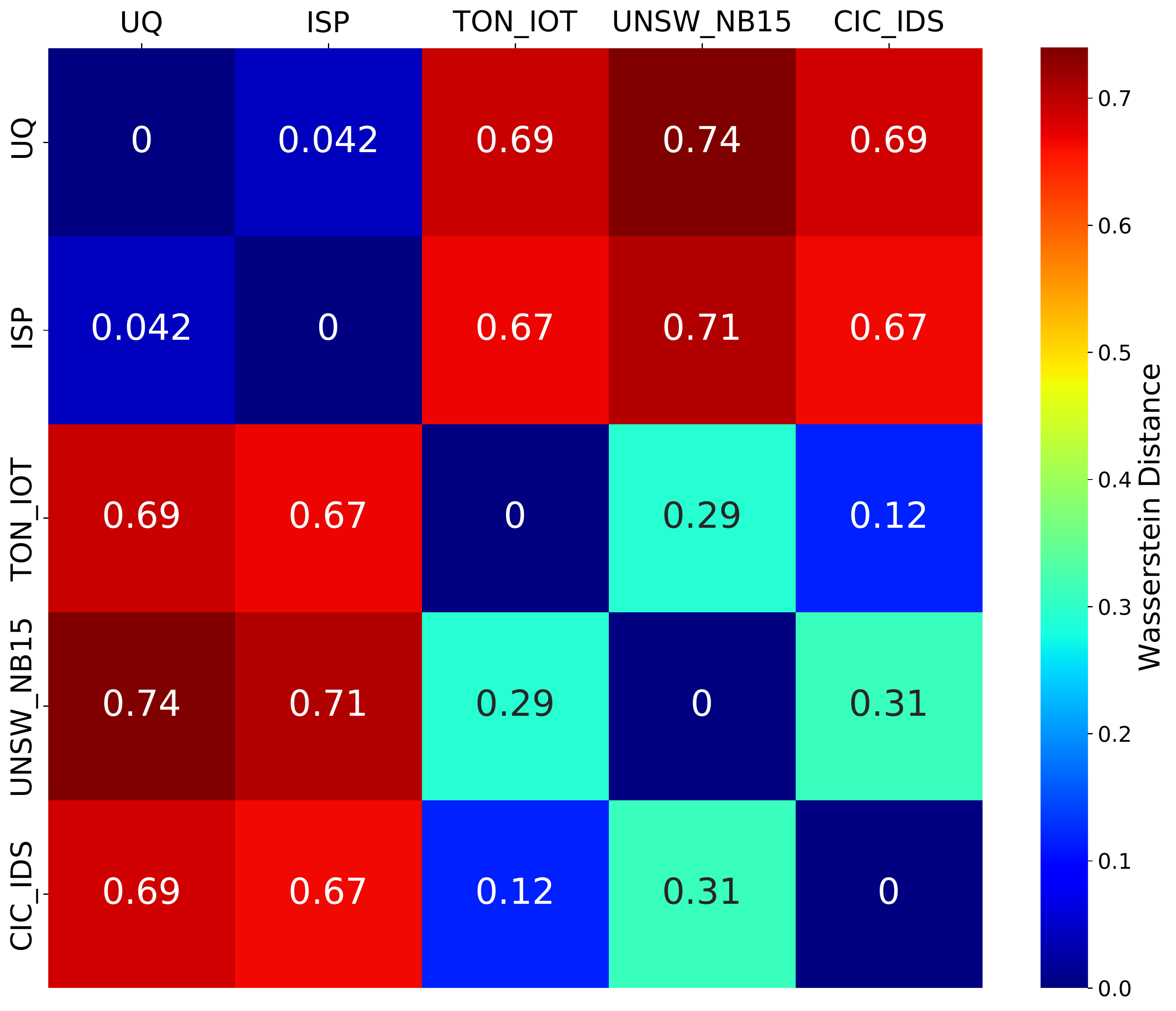}}%
    
    \caption{Wasserstein distances between the samples of the five datasets in terms of a) flow duration, b) flow size, c) (average) packet time, d) packet size, e) number of destination ports per source port, and f) number of  destination IP addresses per source port}%
    \label{Wasserstein distance}
\end{figure}

In this section, we are quantifying the difference between these distributions. There are a range of metrics that can be used to measure the distance between two distributions / probability density functions as reviewed in~\cite{Sung-HyukCha2007}. Each metrics comes with its assumptions about distributions, which defines the accuracy of the measurement.
The distance metric that we are using in this section is \textit{Wasserstein distance}, which is 
also known as the \textit{Earth Mover’s distance}, and is commonly used in the machine learning applications. 
The Wasserstein distance $W$ of two distribution $u$ and $v$ is given by~\cite{Ramdas2017}

\begin{equation}
W(u,v) = \inf_{\pi \in \Gamma(u,v) } \int_{\mathbb{R} \times \mathbb{R} } |x-y| d\pi(x,y)
\end{equation}
\\
\noindent where $\Gamma(u,v)$ is the set of (probability) distributions on $\mathbb{R} \times \mathbb{R}$ whose marginals are $u$ and $v$ on the first and second factors respectively.
The \textbf{inf} stands for \textit{infimum}, also known as the greatest lower bound. If $S$ is a subset of set $T$ (partially ordered), then \textbf{inf}$(S)$ is the greatest element of $T$ which is less than or equal to all elements of $S$.
Th Wasserstein distance of $u$ and $v$ can also be stated in terms of their CDFs, $U$ and $V$ as~\cite{Ramdas2017} 

\begin{equation}
W(u,v) = \int_{-\infty}^{+\infty} |U-V| 
\end{equation}
\\
\noindent which in the case of our study can be applied to CDFs of features computed in Section \ref{Basic Statistical Analysis}.

Figure~\ref{Wasserstein distance} shows the Wasserstein distances between each pair of the five datasets that are used in this study, in the form of a heatmap diagram, for six features.  In Figure~\ref{Wasserstein distance}-a the distance between flow duration distributions is measured. The value of the Wasserstein distances are shown on each entry, in addition to visualising by color. Since the Wasserstein distance metric is symmetric, distance from distribution \textbf{a} to \textbf{b} is the same as distance from \textbf{b} to \textbf{a}. For instance, the distance between UNSW_NB15 and UQ and vice versa is $0.51$. As seen, except CIC-IDS, the distances between synthetic datasets and the real-world datasets is bigger than real-world to real-world distances.

Figure~\ref{Wasserstein distance}-b, c, d, e, and f show the Wasserstein distances for the distribution of flow size, packet time, packet size, number of destination ports per source ports and number of destination IPs per source port, respectively. 
The differences between the real-world and all synthetic datasets are more significant for the rest of features.

\begin{figure}[!b]
    \centering
    \subfloat[\centering ]
    {\includegraphics[width=0.4\columnwidth, height=5cm]{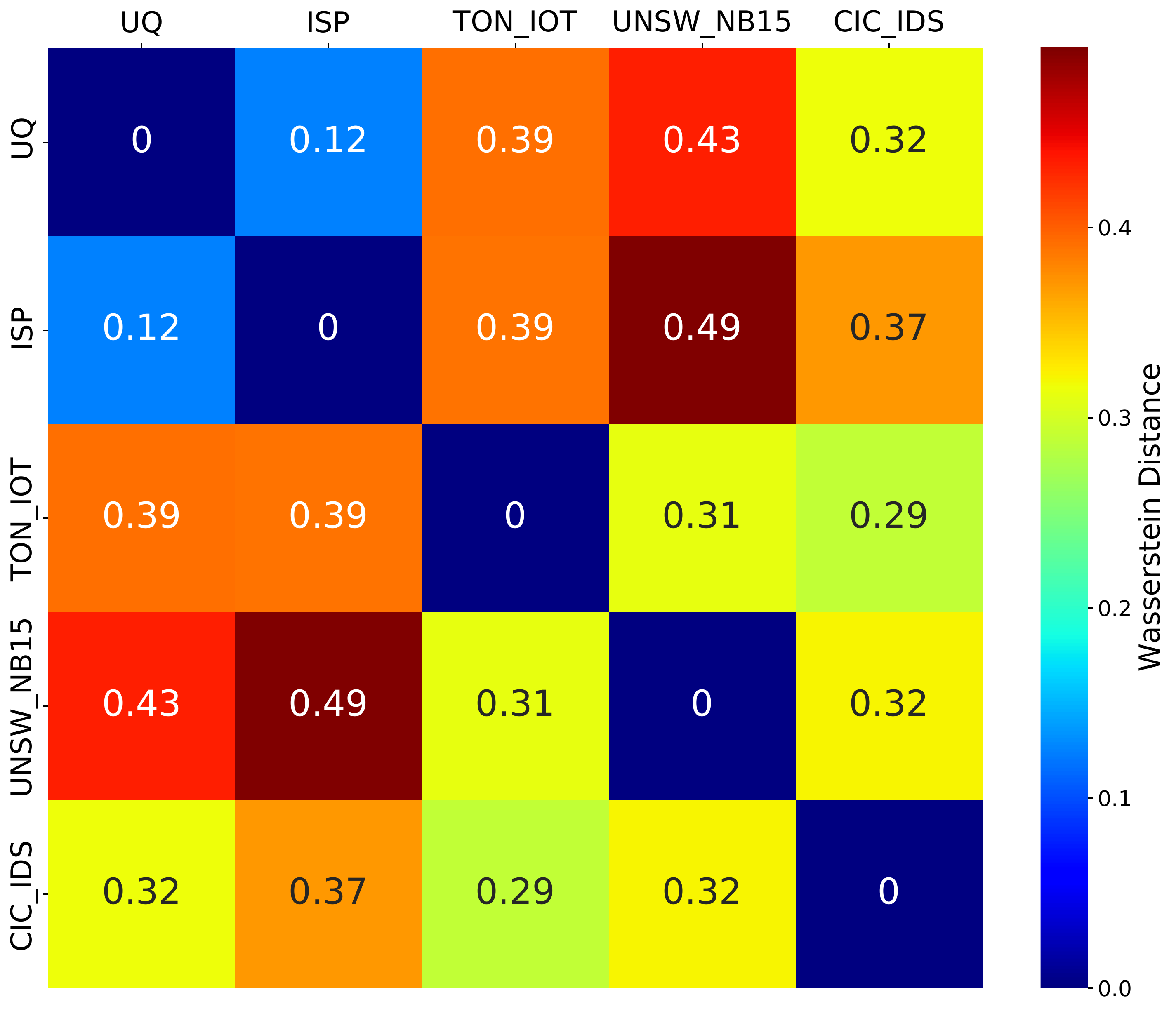} }%
    \hspace{1cm}
        \subfloat[\centering ]
    {\includegraphics[width=0.4\columnwidth, height=5cm]{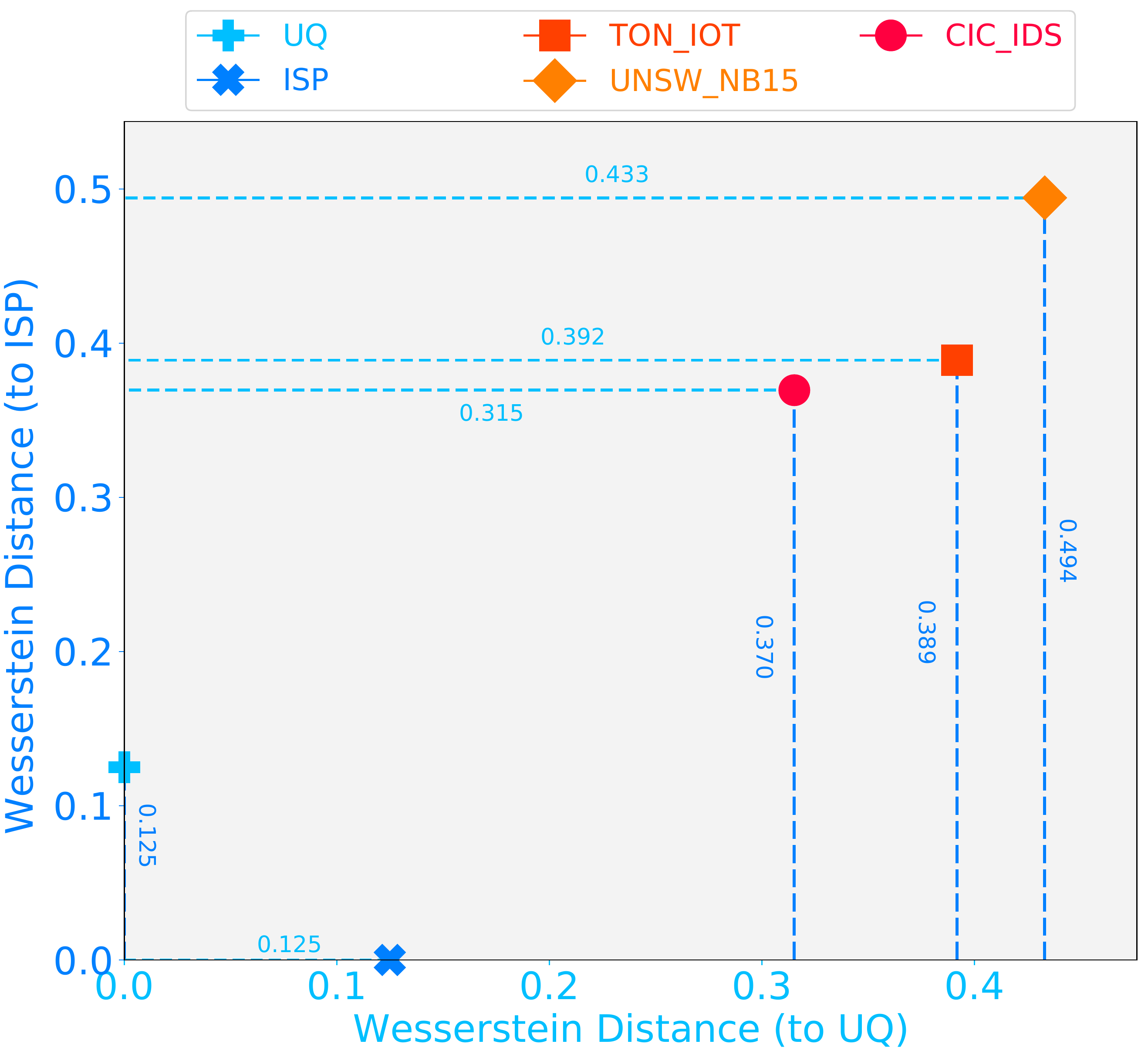} }%

    \caption{Averaged distances between datasets over all 9 features in Table \ref{tab: features} using a) heatmap , and b) distance to UQ and ISP}%
    \label{summary WD}
\end{figure}

In order to summarise the overall differences between these datasets in a single number, the 
Wasserstein distances of all features are calculated and averaged. Figure~\ref{summary WD}-a shows the averaged Wasserstein distance over all features listed in Table \ref{tab: features}. Since the main subject in this paper is the distance between the real-world and synthetic datasets, these values are summarised in Figure~\ref{summary WD}-b. The horizontal axis indicates the distance to UQ dataset, and the vertical axis indicates distance to ISP dataset. As distance of UQ to UQ dataset is zero, it is placed on vertical axis $(x=0)$, similarly the ISP dataset is placed on horizontal axis $(y=0)$. In this way, the coordinates of each dataset represents its averaged distances to the two real-world datasets. As seen, the CIC-IDS dataset is closest dataset to UQ and ISP, and TON_IOT and UNSW_NB15 are placed in farther locations. The main purpose of this representation is to clearly show the distances between the real-world and synthetic datasets. While the two real-world datasets are placed in small distances from each other, all the three synthetic datasets are in much farther distances.

\section{Related Works}\label{Related Works}
In order to select the previous works regarding the subject, we have considered three aspects. First the role of network traffic characteristics, i.e features of network traffic that affect normal network behavior. Then, we looked for those features that are considered for detecting and identifying network traffic anomalies. Finally, we looked for other works in the field of IDS dataset evaluation and summarised these works method and how they evaluated the publicly available IDS datasets.

\subsection{Network Traffic Characteristics}
Many previous works that discuss characteristics of network traffic such as~\cite{KevinThompson1997} and~\cite{Liu2015} focus on traffic volume variations in time to explain main features of network traffic. 
In  \cite{Lakhina2004a}, they use Principle Component Analysis (PCA) to investigate the the origin-destination flows of network as an essential part of network traffic modelling, and finding solution to a wide variety of problems including traffic engineering, capacity planning and anomaly detection. 
While timeseries and time variations of traffic volume play a significant role in many aspects of network such as design and implementation, there are other features of network traffic that are equally important when discussing traffic characteristics. This has been clearly illustrated in two other previous studies~\cite{Kandula2009} and~\cite{Benson2010}  which investigate network traffic not only via its timeseries, but also by analysing its statistical distributions. 
In~\cite{Kandula2009} by collecting a petabyte of measurement data over two months, they discovered and reported traffic characteristics patterns. They not only investigated the features related to traffic volume, but they also studied other features such as flow duration and flow inter-arrival time.

In\cite{Benson2010}, they studied the traffic characteristics of 10 distinct datacentre networks with different organisational administrations such as university, enterprise, and cloud services providers. They studied the traffic patterns of these datacentres by investigating the flow and packet-level properties of various layer-7 applications, and the impact of these applications on network congestion and link and network utilisation. In their study, they investigated a range of packet and flow statistical distributions such as number of flows per second, flow inter-arrival time, flow size, flow duration, and packet size. Unlike previous works, they not only took into account the time domain distribution of the traffic features, but they also considered the statistical measures such as \textit{Cumulative Distribution Function (CDF)}. In addition, they investigated these measures per layer 7 applications, and provided the distribution of corresponding statistical measures per various layer 7 applications. They use these measures to understand the normal patterns of network traffic in various levels such as edge, aggregation and core links.

\subsection{Network Anomalies Characteristics} \label{NAC}
The next group of studies investigate the network traffic characteristics to understand abnormal network behavior and detect anomalies. In this group of studies, various statistics, measures and distributions of several network traffic features have been utilised for detecting abnormal network behaviors. They not only take into account a broader range of network traffic features, they also consider the statistical measures and distributions of these features. Furthermore, some of the studies in this group apply their method on traffic records collected from the real world, mainly backbone networks, which is a stronger evaluation of their proposed method.

In~\cite{Lakhina}, this argument is explained by illustrating the role and importance of the distributions of packet features, such as IP addresses and ports observed in flow traces, in detecting and identifying the structure of a wide range of network anomalies. They state that clustering network traffic based on the distribution of these features creates meaningful clusters of anomalies and such clusters can be utilised for detecting new anomalies and automatic classification of anomalies. In this way, by investigating these distributions, they are able to not only detect the volume-based anomalies, but also detect many other anomalies that do not change traffic volume significantly. 
They validate their proposed methods on data collected from two backbone networks (Abilene and Geant) and conclude that feature distributions is a key ingredient in general network anomaly detection framework.

In~\cite{Soule2007}, they monitored anomalies and collected network traffic records of the same backbone networks (Abilene and Geant), in order to determine which network parameters affect detectability of network anomalies, and how these parameters influence the characteristics of anomalies. They recorded 3 weeks of traffic and routing data from both networks and detected three specific anomalies by applying Kalman filter. 
While they stated that detecting anomalies is significantly dependent on the network design, monitoring infrastructure, and anomaly-detection technique, they investigated variations of the entropy of four features of network traffic, namely source and destination IP addresses and source and destination (L4) Ports.
They concluded that it is not possible to detect all anomalies in a network, based on a single method for Internet-Wide anomaly detection. 

In the last study of this group, based on the idea proposed in~\cite{Lakhina}, using the traffic feature distribution for anomaly detection, histograms of eight network traffic features have been investigated~\cite{histo_anomaly}. These features include the source and destination IP addresses, source and destination (L4) ports, TCP flags, (L4) protocol number, packet size, and flow duration. While they have listed the possible benefits of each feature for detecting specific type of anomaly, they stated that combination of features can reveal changes in the network traffic, which would be invisible otherwise. They applied their proposed method on the data collected from one datacentre, and one campus network and an IDS benchmark dataset. 

The results of these studies, collectively, show that network traffic features distribution play a significant role in the detection of network anomalies in both real world and benchmark datasets. However, this has rarely been investigated in the evaluation of publicly available datasets.
In the next subsection, we briefly explore the available review papers that have evaluated these benchmark datasets.

\subsection{IDS Dataset Evaluation}
While there are many publicly available IDS benchmark datasets that have been frequently utilised for the evaluation of new and existing IDS algorithms, studies that evaluate these datasets themselves are very rare. In this subsection we explore the two studies that we have found in this space. 

The first study investigates usefulness of DARPA dataset for the evaluation of IDSs ~\cite{Thomas2008}. They use two signature-based IDSs, Snort~\cite{snort} and Cisco IDS, along with two anomaly detection methods to evaluate DARPA datasets based on the methodology proposed by MIT Lincoln Laboratory for IDS evaluation. Their results indicate that this dataset is useful for the evaluation of intrusion detection systems. Since this evaluation includes only a single dataset, and it is rather an old study investigating even an older dataset, the results cannot be generalised for other datasets.

The report in \cite{Gharib2017} is the only study we found evaluating multiple IDS datasets. They systematically evaluate 11 publicly accessible IDS datasets published between 1998 till 2016, and conclude that most of these datasets are out of date and unreliable for the evaluation of the IDS algorithms.
By investigating the shortcoming of the existing datasets, they provide a framework for the evaluation of IDS datasets. This framework consists of 11 features which a benchmark IDS dataset should possess, and hence define a corresponding score for each of their studied benchmark datasets. 
These features include complete network configuration, complete traffic, labeled dataset, complete interaction, complete capture, available protocols, attack diversity, anonymity, heterogeneity, feature set, and metadata. 
The main approach in defining these features, which are proposed based on the observations of the existing benchmark datasets, is to prevent the previous dataset shortcomings. However, it is not discussed how having these features guarantees the fitness or enhances the fitness of a synthetic IDS dataset for the real-world evaluation scenarios.

As such, we are proposing for the first time, a criteria for a benchmark dataset, that can be used to evaluate the similarity of a synthetic / testbed-based IDS dataset to the real-world network traffic records. We explain this methodology in the coming sections by first explaining the datasets utilised in our study.

\section{Conclusion}\label{Conclusion}


The benchmark datasets for Network Intrusion Detection Systems (NIDSs) are commonly used by academic and industrial network security researchers for the evaluation of the anomaly detection and NIDS algorithms. In most of the methods based on the machine learning techniques, statistical features of the network traffic play a significant role in the performance of the classification of normal and anomalous traffic. However, these benchmark datasets have never been analysed in terms of the statistical features of their traffic records. The statistical distributions of the attack / anomaly traffic features are usually different from the normal traffic. As such, the normal / benign traffic records of these datasets are the main parts which should simulate / mimic normal / benign the real-world traffic in which the anomaly detection algorithms are supposed to work.

The main purpose of this paper is to introduce tools and methodologies that can measure how realistic is the evaluation of an NIDS algorithm via a synthetic dataset. Currently, this has not been addressed in the network security research field. This paper, tries to address this gap not only by proposing the required tool and methodology, but also by applying the proposed methodology on three recent NIDS datasets, and comparing them to two diverse real-world network traffic datasets.

We initially propose nine traffic features that can be used for this purpose. Then, we illustrate the statistical distributions of these features for
the synthetic and real-world datasets. We also use four different dimensionality reduction methods to embed the set of nine features in a 2-dimensional Euclidean space and visualise the resulting embeddings for all datasets. In both cases, the statistical distributions and dimensionality reduced visualisations, there are significant distinctions between the two groups of datasets, the real world and synthetic. In addition, the differences among members of each group is much less significant, e.g. the two real-world datasets are more similar together than to the synthetic datasets, despite their diverse origin.

While the illustrated distributions clearly indicate the differences in the traffic features of the synthetic and real-world datasets, differences between the real-world datasets and each of the synthetic datasets varies from one dataset to another. We propose a metric to quantify these differences between statistical distributions of features of each pair of datasets. The proposed metric also clearly indicates that difference between the two real-world datasets, despite the fact that they represent two different types of networks from two different organisation and geographical-location, is much less than to either of synthetic datasets. 

Based on these results, we believe that the evaluation of anomaly detection and NIDS algorithms using synthetic datasets, which are not created in the context of real-world traffic, does not guarantee their classification performance in the real-world scenarios. The addressing of this gap constitutes the future work of our research where we are proposing a methodology for creating NIDS benchmark datasets that their traffic feature distributions comply with that of the real-world traffic.

\bibliography{main}
\bibliographystyle{ieeetr}

\end{document}